\documentclass[aps,prd,twocolumn,showpacs,preprintnumbers,amsmath,amssymb]{revtex4}

\usepackage{amsmath,amssymb}

\usepackage{color}
\usepackage[dvips]{graphicx}
\usepackage{subfigure}

\newcommand{\lsim}{\mathrel{\mathop{\kern 0pt \rlap
  {\raise.2ex\hbox{$<$}}}
  \lower.9ex\hbox{\kern-.190em $\sim$}}}
\newcommand{\gsim}{\mathrel{\mathop{\kern 0pt \rlap
  {\raise.2ex\hbox{$>$}}}
  \lower.9ex\hbox{\kern-.190em $\sim$}}}

\newcommand{\sigmav}{\langle \sigma_{\rm ann} v \rangle}
\newcommand{\sigmavzero}{\langle \sigma_{\rm ann} v \rangle_0}

\newcommand{\beq}{\begin{equation}}
\newcommand{\eeq}{\end{equation}}
\newcommand{\bea}{\begin{eqnarray}}
\newcommand{\ena}{\end{eqnarray}}
\newcommand{\etal}{{\it et al.}}

\newcommand{\into}{\rightarrow}
        
\newcommand{\be}{\begin{equation}}
\newcommand{\ee}{\end{equation}}
\newcommand{\ba}{\begin{array}}
\newcommand{\ea}{\end{array}}

\newcommand{\eea}{\end{eqnarray}}

\begin{document}

\preprint{DFTT 08/2006}
\preprint{DESY 06 -- 077}

\title{Constraining pre Big--Bang--Nucleosynthesis Expansion using
Cosmic Antiprotons}
%\title{Constraints on Dark Energy Models from Cosmic Antiprotons}



%
\author{Mia Schelke}
%\email{schelke@to.infn.it}
%\homepage{http://www.astroparticle.to.infn.it}
\affiliation{Istituto Nazionale di Fisica Nucleare, Sezione di Torino
\\ via P. Giuria 1, I--10125 Torino, Italy \\ {\tt (schelke@to.infn.it)}}

\author{Riccardo Catena} 
\affiliation{Deutsches Elektronen-Syncrotron DESY, \\ 22603 Hamburg, Germany \\ {\tt (catena@mail.desy.de)}}

\author{Nicolao Fornengo}
%\email{fornengo@to.infn.it}
%\homepage{http://www.astroparticle.to.infn.it}
%\homepage{http://www.to.infn.it/~fornengo}
\affiliation{Dipartimento di Fisica Teorica, Universit\`a di Torino \\
Istituto Nazionale di Fisica Nucleare, Sezione di Torino \\
via P. Giuria 1, I--10125 Torino, Italy \\ {\tt (fornengo@to.infn.it)}}

\author{Antonio Masiero}
\affiliation{Dipartimento di Fisica, Universit\`a di Padova, \\
Istituto Nazionale di Fisica Nucleare, Sezione di Padova \\ via
Marzolo 8, I-35131, Padova, Italy \\ {\tt (masiero@pd.infn.it)}}

\author{Massimo Pietroni} 
\affiliation{Istituto Nazionale di Fisica Nucleare, Sezione di Padova \\ via
Marzolo 8, I-35131, Padova, Italy \\ {\tt (pietroni@pd.infn.it)}}

\date{\today}

\begin{abstract}
A host of dark energy models and non--standard cosmologies predict an
enhanced Hubble rate in the early Universe: perfectly viable models,
which satisfy Big Bang Nucleosynthesis (BBN), cosmic microwave
background and general relativity tests, may nevertheless lead to
enhancements of the Hubble rate up to many orders of magnitude. In
this paper we show that strong bounds on the pre--BBN evolution of the
Universe may be derived, under the assumption that dark matter is a
thermal relic, by combining the dark matter relic density bound with
constraints coming from the production of cosmic--ray antiprotons by
dark matter annihilation in the Galaxy. The limits we derive can be
sizable and apply to the Hubble rate around the temperature of dark
matter decoupling. For dark matter masses lighter than 100 GeV, the
bound on the Hubble--rate enhancement ranges from a factor of a few to
a factor of 30, depending on the actual cosmological model, while for
a mass of 500 GeV the bound falls in the range 50--500. Uncertainties
in the derivation of the bounds and situations where the bounds become
looser are discussed. We finally discuss how these limits apply to
some specific realizations of non--standard cosmologies: a
scalar--tensor gravity model, kination models and a Randall--Sundrum
D--brane model.
\end{abstract}

\pacs{95.35.+d,95.36.+x,98.80.-k,04.50.+h,96.50.S-,98.70.Sa,98.80.Cq}
% 95.35.+d Dark matter
% 95.36.+x Dark energy
% 98.80.-k Cosmology
% 04.50.+h Gravity in more than four dimensions, KK theory, unified
%          field theories, alternatives theories of gravity
% 96.50.S- Cosmic rays in the solar system
% 98.70.Sa Cosmic rays
% 98.80.Cq Particle theory and field theory models in the early Universe

\maketitle

%%%%%%%%%%%%%%%%%%%%%%%%%%%%%%%%%%%%%%%%%%%%%%%%%%%%%%%%%%%%%%%%%%%%%%%%%
\section{Introduction}
\label{sec:intro}

Current cosmological and astrophysical observations clearly show that
our Universe is dominated by two unknown and exotic components, dark
matter and dark ener\-gy, whose energy densities are measured to fall in
the following ranges (at $2 \sigma$ C.L.)  \cite{omegalimits}:
\be
0.092\le\Omega_{\rm CDM} h^2 \le 0.124
\label{oh2 constraint}
\ee
for the cold dark matter (CDM) component, responsible of structure formations
and galactic and extragalactic dynamics, and:
\be
0.30 \le\Omega_{\rm DE} h^2 \le 0.46
\label{eq:de}
\ee
for the unclustered dark energy component, which is responsible of the
current accelerated expansion of the Universe. (As usual $\Omega$ denotes the 
ratio between the mean density and the critical density and $h$ is the Hubble 
constant in units of $100\ \textrm{km}\ \textrm{s}^{-1}\ \textrm{Mpc}^{-1}$.) 

The nature of both components is unknown. A common and appealing
possibility is that dark matter is composed by elementary particles
which decoupled from the thermal plasma in the early Universe. The
dark energy component poses more serious problems: a possibility is
that it is due to the presence of a scalar field, whose cosmological
dynamics allows it to become the dominant component of the Universe
just recently in the evolu\-ti\-o\-nary history of the Universe. Most of the
dark energy models predict that the expansion rate of the Universe may 
have been different from the one
predicted by the standard Friedman--Robertson--Walker (FRW) model at 
very early stages. Not only dark energy models, but also other 
cosmological mo\-dels, can predict an enhancement of the Hubble rate 
\cite{note on reduction}. 
If this occurs around the time when the
dark matter particles decouple from the thermal bath, the change in
the Hubble rate may leave its imprint on the relic abundance of the
dark matter. This has been discussed in details in dark energy models
based on scalar--tensor gravity \cite{scalartensorrelic} and in 
quintessence models with a kination phase 
\cite{Salati,Rosati,profumoullio,Pallis} and for anisotropic 
expansion and other models of modified expansion 
\cite{Barrow,kamionkowskiturner}. This effect implies that
the basic CDM properties may depend on the specific cosmological 
model. Thus, information on the CDM particles may be used to constrain
cosmological models and vice versa. In this paper we will exploit this
connection between the dark matter and the expansion rate in order to
derive bounds on cosmological models from observational data related to
the dark matter particles. We emphasize that the effects we are going
to study do not arise because of a direct coupling between dark
energy and dark matter: they are instead due to the effect induced
by the dark energy model (or other models with enhanced expansion) 
on the decoupling of the dark matter particle.

We will consider a generic Weakly Interacting Massive Particle (WIMP)
as candidate of cold dark matter. For the cosmological model we
consider models that lead to an enhancement of the Hubble expansion
rate in the early Universe as compared to the rate in standard 
cosmo\-lo\-gy, like those in
Refs. \cite{scalartensorrelic,Salati,Randall}. In order to be general
in our analysis, we will consider a suitable parametrization of the
enhancement of the Hubble rate in the early Universe. 
%This parametrization may be applied to diffe\-rent classes of dark energy
%models. 
As specific examples we will then relate our parametrization
to scalar--tensor gravity (ST) models \cite{scalartensorrelic}, to
models with a kination phase \cite{Salati} and to a Randall--Sundrum
D--brane model \cite{Randall}.

The enhancement of the Hubble expansion rate in the early Universe can
affect phenomena that are sensitive to the exact time at which they
occur. This is the case for the freeze--out of a WIMP. The enhanced
Hubble rate causes an enhanced WIMP dilution. The WIMP annihilation
rate therefore cannot keep up with the expansion as long as in the
standard case. Consequently, the WIMP freezes out earlier in this kind
of non--standard cosmologies, leading to a higher relic abundance.
The enhancement of the WIMP relic abundance can be very dramatic, up
to a few orders of magnitude \cite{scalartensorrelic,Salati}. The
requirement that the WIMP is the dominant dark matter component, {\em
i.e.}~that its relic abundance satisfies the bound of Eq.~(\ref{oh2
constraint}), implies that the properties of the successful WIMP are
dramatically changed with respect to the standard FRW case. Since the
dark matter relic abundance, apart from the expansion, mainly depends 
on the WIMP mass and
annihilation cross section, we can say that the cosmolo\-gical model has left
its fingerprint on the dark matter. A fingerprint which could turn out
to be one of the most important signatures of these dark energy and other cosmological models.

In fact the WIMPs that in the modified scenario have the correct relic
density possess a higher annihilation cross section than do WIMPs
selected by standard cosmology. This means that indirect detection
signals are favoured in the modified scenario. In particular, it has
been shown that the antiproton component in cosmic rays is a power\-full
tool to constrain dark matter properties, even though it is affected
by large astrophysical uncertainties \cite{pb}. In Ref.~\cite{pbarlim}
combined limits in the astrophysical--WIMP parameter space have been
derived and it has been shown that antiprotons can power\-fully set 
bounds on WIMPs
in the tens--hundreds of GeV range. We are going to apply the same
argument here, from which we will be able to derive constraints on the
cosmological models with enhanced Hubble expansion. We will show that
the constraints on dark matter coming from antiproton indirect
searches strongly constrain the cosmological models with enhanced
Hubble rate in the early Universe. When one studies a specific cosmological 
model, the model can be constrained by a number of relevant
observables, related to {\em e.g.}~Big Bang Nucleosynthesis (BBN) 
\cite{Cyburt},
Cosmic Microwave Background (CMB), Large Scale structure and
Supernovae \cite{Wang, Zhao}, weak lensing \cite{Albert} and General
Relativity tests \cite{Cassini}. We will show that the antiproton
bound, under suitable condition, may result in even stronger limits.

In Section \ref{sec:hubble} we discuss our parametrization of the
enhanced Hubble rate, and in Section \ref{sec:cosm models ex} we show
its connection to specific cosmological models. In Section
\ref{sec:relic} we discuss the relic density calculation 
in modified
cosmology, for which some useful analytical approximations are given
in Appendix \ref{app:analytic}.  Section \ref{sec:pbar} deals
with the cosmic antiproton signal and the bounds on the dark matter. 
In Section \ref{sec:constraints} we then derive the
constraints on the cosmolo\-gi\-cal models. Section \ref{sec:specific}
translates our bounds to some specific cosmological models, namely
scalar--tensor cosmology, kination models and Randal--Sundrum D--brane
cosmology. Finally, in Section \ref{sec:conclusion} we summarize
our main results.

%%%%%%%%%%%%%%%%%%%%%%%%%%%%%%%%%%%%%%%%%%%%%%%%%%%%%%%%%%%%%%%%%%%%%%%%%
\section{The Hubble enhancement function}
\label{sec:hubble}

\begin{figure}[t] \centering
\vspace{-20pt}
\includegraphics[width=1.0\columnwidth]{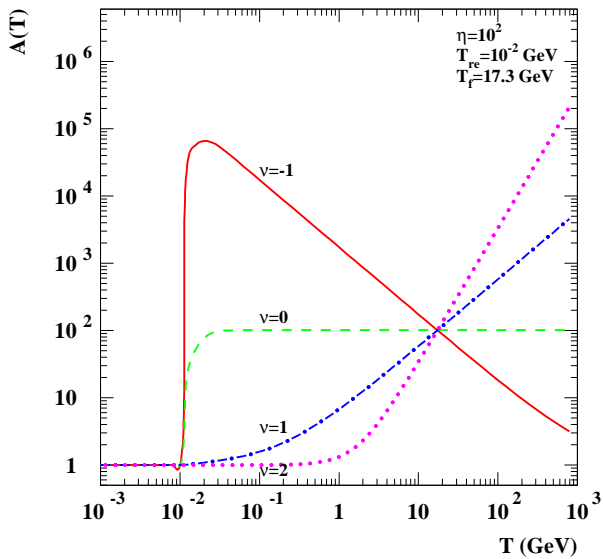}
\vspace{-20pt}
  \caption{The Hubble rate enhancement function $A(T)$ as a function
  of the temperature. Notice that time is running from right to
  left. The solid red line has a slope parameter $\nu = -1$, the
  dashed green line $\nu = 0 $, the dash-dotted blue $\nu = 1$ and the
  dotted purple line has $\nu = 2$. For all the curves we have chosen
  $\eta = 10^2$ and $T_\textrm{re} = 10^{-2}$ GeV. The freeze out
  temperature of the relic WIMP has been fixed to $T_\textrm{f} =
  17.3$ GeV (a case which refers, {\rm e.g.}, to a particle with mass
  of 500 GeV and annihilation cross section
  $\langle\sigma_{\textrm{ann}}v\rangle =
  10^{-7}\,\textrm{GeV}^{-2}$).}
\label{fig:fnew01}
\end{figure}

Let us consider a class of cosmological models that posses a Hubble
rate $H$ in the early Universe enhanced with respect to its value in
standard cosmology $H_{\textrm{std}}$. We parametrize this enhancement by
means of a function $A(T)$:
\bea
\label{h.Eq.a.h.gr}
H&=&A(T)H_{\textrm{{std}}}\qquad\qquad\textrm{at early times}\\
H&=&H_{\textrm{{std}}}\qquad\qquad\qquad\, \textrm{at later~times} 
\eea
This situation occurs {\em e.g.} in scalar-tensor gravity models and
in models with a kination phase. Also some models with extra
dimensions lead to an enhanced Hubble rate. We will review some of
these examples in the next Section to show that they can all be
covered by the following parametrization of the enhancement function
$A(T)$:
\be
A(T)=1+\eta\left(\frac{T}{T_\textrm{f}}\right)^\nu
  \tanh\left(
 \frac{T-T_{\textrm{re}}}{T_{\textrm{re}}}\right)
\label{A def}
\ee
for temperatures $T>T_\textrm{re}$ and $A(T)=1$ for $T\leq
T_\textrm{re}$. By $T_{\textrm{re}}$ we denote the temperature at
which the Hubble rate ``\emph{re--enters}'' the standard rate and by
$T_f$ a reference temperature, which we specify below. Clearly the
Hubble parameter has to approach the General Rela\-ti\-vi\-ty (GR) case at
some epoch $T_\textrm{re} \gtrsim 1\ \textrm{MeV}$, in order not to spoil 
the successful predictions of Big
Bang Nucleosynthesis and the formation of the Cosmic Microwave
Background Radiation (CMB). These two key events in the history of the 
Universe are typically among the major constraints on DE theories.

In fig.~\ref{fig:fnew01} we show $A(T)$ for $\eta=100$,
$T_\textrm{re}=10^{-2}$ GeV and four different values of the exponent
$\nu$, which we will use throughout the paper as reference points in
our ana\-ly\-sis. The figure shows that at the temperature $T_\textrm{f}$
(which here is $T_f \gg T_{\textrm{re}}$) all curves take the same
value $\eta$ (since $\eta \gg 1$). In our analysis, we will take
$T_\textrm{f}$ as the temperature at which the WIMP freezes out in 
standard cosmology. In a situation like the one shown in the figure,
the parameter $\eta$ gets the meaning of enhancement factor of the
Hubble rate at the time of WIMP freeze--out.

The slope of the curve around $T_\textrm{f}$ is mostly determined by
the exponent $\nu$. We arbitrarily use in Eq. (\ref{A def}) a hyperbolic tangent
to assure that all curves approach unity as a continuos function when
the temperature of the Universe approaches $T_{\textrm{re}}$. However,
for analytic approximations that may help to understand our numerical
results, a sharp jump to the standard case at $T=T_{\textrm{re}}$ is
appropriate. Changes in the slope of the re--entering phase will be
briefly discussed in Sect. \ref{subsec:relax assumptions}.

%%%%%%%%%%%%%%%%%%%%%%%%%%%%%%%%%%%%%%%%%%%%%%%%%%%%%%%%%%%%%%%%%%%%%%%%%
\section{Cosmological models with a modified expansion rate}
\label{sec:cosm models ex}

In this section we present a short list of interesting cosmological
models which lead to an enhanced  expansion rate  as compared to the 
standard Friedman--Robertson--Walker (FRW) model based on General
Relativity.

\subsection{Scalar--tensor theories}\label{subsec:ST 1}

In a scalar--tensor theory of gravity both a metrical tensor
$g_{\mu\nu}$ and a scalar field $\varphi$ are involved in the
description of the gravitational interaction \cite{Damour}. Because of
this additional scalar degree of freedom, the Universe expands
differently compared to a standard FRW solution of the Einstein
gravity equations \cite{scalartensorrelic}.  This class of theories
can be formulated in different frames related to each other by a Weyl
re--scaling of the metric \cite{Esposito}. As pointed out in Ref.
\cite{Dicke}, a frame transformation amounts in a change of units and
therefore the physical results cannot depend on the frame.  A
frame-independent formulation of the scalar--tensor cosmology is given
in \cite{Pietroni}. In order to deal with scalar field-independent
masses and couplings, in the language of scalar--tensor theories this
means that we stay in the Jordan frame. This was our approach in the
calculation of the effects induced on the WIMP decoupling in
scalar--tensor theories in Ref. \cite{scalartensorrelic}.

The ratio between the Jordan frame expansion rate and the standard
General Relativity expansion rate is given by \cite{scalartensorrelic}:
\begin{equation}
\frac{H^2}{H_{std}^2}=A^2(\varphi) \frac{\left[ 1+\alpha(\varphi)
\varphi^ {\prime} \right]^2}{1-(\varphi^ {\prime})^2/6}
\label{ratio}
\end{equation}
where $\alpha(\varphi)=d \log A(\varphi)/d\varphi$, $A(\varphi)$ is
the Weyl factor relating the Jordan and Einstein frame and a prime
denotes derivation with respect to the logarithm of the scale factor.

To find a general correspondence between our pa\-ra\-me\-tri\-zation in
Eq. (\ref{A def}) and a scalar--tensor behaviour of the expansion rate
is quite difficult. This is due to the fact that the scaling of the
ratio in Eq.~(\ref{ratio}) is deeply related with the cosmological
dynamics of the scalar field. However, as shown in a specific example
in Ref.~\cite{scalartensorrelic}, a numerical solution of the scaling
behavior of Eq. (\ref{ratio}) can be translated in terms of our
parametrization. In the simplest case of a slowly varying scalar field
($\varphi^ \prime \simeq 0$) we have $H = A(\varphi)H_\textrm{std}$, and 
in the specific example of \cite{scalartensorrelic} we have $A(\varphi) =
A(\varphi(T_\textrm{f}))(T/T_\textrm{f})^{-0.82}$. In terms of our
parametrization, this example has $\eta = A(\varphi(T_\textrm{f}))$
and $\nu = -0.82$. Fig. 6 of Ref. \cite{scalartensorrelic} shows that
the enhancement of the Hubble rate around the WIMP freeze--out may be
quite sizeable, up to factors of $10^4$. We remind that the model
studied in Ref. \cite{scalartensorrelic} was a perfectly viable
scalar--tensor model, since it evaded all experimental constraints
from BBN, CMB and gravitational probe limits, like the Cassini
mission \cite{Cassini}. The enhanced Hubble rate reflects in an 
anticipated WIMP
decoupling, with an ensuing larger relic abundance. Ref.
\cite{scalartensorrelic} showed that for such a fast re--entering of
the Hubble rate on its GR behaviour, it may be possible that the
already--decoupled WIMPs start a brief phase of {\em
re--annihilation}, due to the fact that they are still over--abundant
and therefore their annihilation rate is larger than the GR Hubble
rate. This phenomenon has the consequence of reducing the WIMP relic
abundance with respect to the value it would have had without
re--annihilation. Nevertheless, the outcome of this scalar--tensor
model is that the WIMP current abundance may be larger that the GR
one by up to 2-3 orders of magnitude. This result would then be even
stronger if a re--annihilation phase does not occur, due to a slower
re--entering of the Hubble rate to GR. Summa\-rizing: perfectly viable
scalar--tensor models from the point of view of cosmological and
astrophysical observations, may predict a strongly enhanced WIMP relic
abundance for a give candidate. This was our original motivation to
try to set additional limits on these cosmologies by means of
observables related to the dark matter sector, namely the antiproton
indirect--detection signal, as it will be described in the next
Sections. However, our discussion may be applied also to other
cosmological scenarios with enhanced early Hubble rate, like the
following two relevant cases.

\subsection{Kination}\label{subsec:kination 1}

A kination is a period in which the total energy density of the
Universe is dominated by the kinetic term of a scalar field. A phase of kination is generically expected in quintessence models based on tracking solutions for the scalar field, Ref.~\cite{Steinhardt}. When the scalar potential 
$V(\Phi)$ is
negligible compared to the kinetic energy of the scalar field, the
total energy density in the scalar field, $\rho_{\Phi}$, scales like
$\sim a^{-6}$ (where $a$ is the scale factor of the Universe). This
means that during kination $H^2 \propto \rho_{\textrm{tot}} \simeq
\rho_{\Phi} \propto a^{-6}$. More precisely, the ratio between the
expansion rate $H$ during a kination period and the standard expansion
rate $H_{std}$ is given by:
\begin{equation}
\frac{H^2}{H^2_{std}}= 1+\frac{\rho_{\Phi}}{\rho_r} \,,
\label{rapk}
\end{equation} 
where $\rho_{\Phi}/\rho_{r}$ can be written as \cite{Salati}
\begin{equation}
\frac{\rho_{\Phi}}{\rho_{r}}= \eta_{\Phi} \left[
\frac{h(T)}{h(T_\textrm{f})} \right]^2 \frac{g(T_\textrm{f})}{g(T)}
\left( \frac{T}{T_\textrm{f}} \right)^2 \simeq \eta_{\Phi} \left(
\frac{T}{T_\textrm{f}} \right)^2
\label{hg}
\end{equation} 
with $\eta_{\Phi}=\rho_{\Phi}(T_\textrm{f})/\rho_{r}(T_\textrm{f})$
and $h_{\textrm{eff}}$ and $g_{\textrm{eff}}$ respectively the
entropy--density and energy--density effective degrees of freedom. The
approximation in Eq. (\ref{hg}) is justified only in a range of
temperatures where $h_{\textrm{eff}}$ and $g_{\textrm{eff}}$ do not
change considerably with respect to their value at $T_\textrm{f}$.

From Eq.~(\ref{rapk}) and (\ref{hg}) we see that a kination model can
be approximated by our parametrization once the following values for
the parameters are chosen: $\nu=1$ and $\eta=\sqrt{\eta_{\Phi}}$.

\subsection{Extra Dimensions}
\label{subsec:extra dim 1}

We refer to the RSII model \cite{Randall}. In this model a single
3--brane is embedded in a five dimensional bulk with a negative five
dimensional cosmological constant. Unlike other Extra Dimension
models, here the fifth dimension is not compact and no orbifold
boundary conditions are imposed. Moreover, the five--dimensional
metric is not factorizable and an exponential function of the fifth
coordinate multiplies the four dimensional metric. With this set up, a
Kaluza Klein (KK) reduction of the five dimensional gravitational
excitations gives rise to a spectrum with a zero mode (the standard
four--dimensional graviton) localized in the extra dimension and a
conti\-nu\-um of massive KK modes weakly coupled to the low energy states
on the brane.  Due to the fact that the four--dimensional gravitons
can propagate in a confined region of the fifth dimension and since at
low energy the massive KK modes are only weakly coupled, the theory is
not in contrast with experimental gravity even if the volume of the
extra dimension is infinite. The Einstein equation on the 3-brane are
studied in Ref. \cite{Shiromizu} and the cosmology of such a model is
carefully investigated in Ref. \cite{Durrer}. In particular it has
been shown that the ratio between the expansion rate $H$ of such a
model and the expansion rate $H_{std}$ of standard cosmo\-lo\-gy is given
by:
\begin{equation}
\frac{H^2}{H^2_{std}}= 1+\frac{\rho}{2\lambda}
\label{rape}
\end{equation} 
where $\lambda$ is the tension on the brane and $\rho$ the energy
density on the brane. When the total energy density $\rho$ equals the
radiation energy density $\rho_r$ we have:
\begin{equation}
\frac{H}{H_{std}}\simeq \sqrt{\frac{\rho_r(T_\textrm{f})}{2 \lambda}}
\left(\frac{T}{T_\textrm{f}}\right)^2
\end{equation}
Comparing this expression with Eq.~(\ref{A def}), we see that such a
model is described by our parametrization with the values of the
parameters $\nu=2$ and $\eta=\sqrt{\rho_r(T_\textrm{f}) /
(2\lambda)}$.

\section{The relic density calculation}
\label{sec:relic}

In this section we will briefly discuss how a cosmolo\-gi\-cal model with
enhanced Hubble rate affects the relic abundance of WIMP dark
matter. This effect was stu\-died in details in
Ref.~\cite{scalartensorrelic} for the scalar--tensor case and 
in Refs.~\cite{Salati,Rosati,profumoullio,Pallis} for the 
kination case. See also Refs.~\cite{Barrow,kamionkowskiturner} 
for anisotropic expansion and other models of modified expansion.

The evolution of the WIMP number density $n$ as a function of
cosmological time $t$ is described by the standard Boltzmann equation,
the only difference being that the standard Hubble parameter
$H_{\textrm{{std}}}(T)$ is now replaced by the modified Hubble rate,
$H(T) = A(T)H_{\textrm{{std}}}(T)$:
\be
 \frac{dn}{dt}=-3Hn-\langle\sigma_{\textrm{ann}} v\rangle 
                       (n^2-n_{\textrm{eq}}^2)
\label{boltzmann eq}      
\ee
where $\langle\sigma_{\textrm{ann}} v\rangle$ is the usual thermally
averaged value of the WIMP annihilation cross section times the
relative velocity. The modification of the Hubble rate can be
rephrased as a change in the effective number of degrees of freedom
from $g_{\textrm{eff}}(T)$ to $A^2(T)g_{\textrm{eff}}(T)$. The Boltzmann 
equation is
more conveniently solved by rewriting it in terms of the comoving
abundance $Y = n/s$ where $s$ is the entropy density $s =
(2\pi^2/45)h(T)T^3$ and studying the evolution as a function of the
temperature $T$:
\begin{equation}
\frac{dY}{dx} = -\frac{1}{x}\frac{s}{H}\langle\sigma_{\textrm{ann}}v\rangle(Y^2-Y^2_\textrm{eq})
\label{eq:boltzmann0}
\end{equation}
where $x = m_\chi/T$. In our analysis we solve the Boltzmann equation
Eq. (\ref{eq:boltzmann0}) numerically down to the current value of the
comoving abundance $Y_0$. The WIMP relic abundance is then simply:
\be
\Omega_\chi h^2 = \frac{m_\chi s_0 Y_0}{\rho_c}
\label{eq:relic}
\ee
where $s_0$ and $\rho_c$ are the current values for the entropy
density and the critical mass--density of the Universe. In order to
get some insight in our numerical results, we report in Appendix
\ref{app:analytic} some useful analytical approximations which are
valid for large enhancements and re--entering temperature (much) lower
than the temperature at which decoupling occurs. Notice that in our
definition of the function $A(T)$ in Eq. (\ref{A def}), we use as a
normalization temperature the freeze--out temperature obtained in standard 
cosmo\-lo\-gy and defined in Appendix \ref{app:analytic}. We report it here for
convenience:
\be
x_\textrm{f} = \ln\left[0.038\,m_\textrm{pl}\,g\,m_\chi 
{\langle\sigma_{\textrm{ann}}v\rangle_{T_\textrm{f}}x_\textrm{f}^{-1/2}}
{g^{-1/2}_{\textrm{eff}}(x_\textrm{f})}\right]
\label{eq:xf}
\ee

The effect of the enhanced Hubble rate on the WIMP relic density can
be very important. As was found in Ref. \cite{scalartensorrelic}, and
as we are also going to see later in this paper, the relic density can
be up to few orders of magnitude larger in the modified scenario as
compared to the standard case. From combined cosmological observations
we have the very stringent bound of Eq. (\ref{oh2 constraint}) for the
cold dark matter density. This means that the WIMP candidates selected in the
case of enhanced Hubble rate will be vastly different ({\em i.e.}~have
different values of their relevant parameters, like mass and
couplings) from the WIMPs that fit into the standard cosmology
picture. The cosmolo\-gi\-cal model has, in other words, left its signature
on the dark matter. A signature which might turn out to give valuable
clues.

Not only can we say that the WIMPs selected in the standard and
modified cases are different, we are also able to say something
general about the difference of their phenomenology. This builds on
the fact that the standard WIMP relic abundance in general is
approximately inversely proportional to the WIMP annihilation cross
section. Analytically one finds (see Appendix \ref{app:analytic}) the
well--known behaviour:
\be
\Omega_\chi h^2 \sim \frac{1}{\langle\sigma_{\rm ann}v\rangle_{\rm int}}
\label{oh2 approx}
\ee
where $\langle\sigma_{\rm ann}v\rangle_{\rm int}$ is the following
integration of the thermal annihilation cross--section:
\begin{equation}
\langle\sigma_{\rm ann}v\rangle_{\rm int} \equiv \frac{1}{{\cal
G}(x_{\rm f})}\int_{x_{\rm f}}^\infty \frac{{\cal
G}(x)\,\langle\sigma_{\textrm{ann}}v\rangle}{A(x)\, x^2}\, dx
\label{eq:sigmaint}
\end{equation}
The WIMPs that fulfill the relic density constraint of Eq.~(\ref{oh2
constraint}) when calculated with the enhanced Hubble rate, would have
a much lower density when recalculated in standard cosmology. From
Eqs.~(\ref{oh2 approx},\ref{eq:sigmaint}) this then means that these
WIMPs have a higher annihilation cross section than WIMPs that satisfy
the relic density constraints when calculated with the standard Hubble
expansion. A high WIMP annihilation cross section in the early
Universe does in general mean that also the current WIMP annihilation
rate in the Galaxy is high
\footnote{This is the case when the cross section does not exhibit a
strong temperature--dependence and when thermal effects like
coannihilations are not effective in the determination of the thermal
average of the cross section. This is the case we shall focus on
mainly in this paper. In Sec.\ref{subsec:relax assumptions} we study a
temperature dependent cross section.}.
The WIMP candidates selected by the models of enhanced expansion rate
are therefore in general more suitable for indirect detection than are
the WIMP candidates selected in the standard scenario. At the same
time this means that the bounds on the WIMP annihilation cross section
coming from searches for indirect WIMP signals can be used to
constrain the expansion rate in the early Universe. In this paper we are 
going to
analyze the constraints coming from the cosmic antiproton signal,
which is the only indirect probe for which strong constraints can be
determined \cite{pbarlim}.

%%%%%%%%%%%%%%%%%%%%%%%%%%%%%%%%%%%%%%%%%%%%%%%%%%%%%%%%%%%%%%%%%%%%%%%%%
\section{The cosmic antiproton signal and the bound on the annihilation cross section}
\label{sec:pbar}

\begin{figure}[t]
\vspace{-27pt}
\includegraphics[width=1.10\columnwidth]{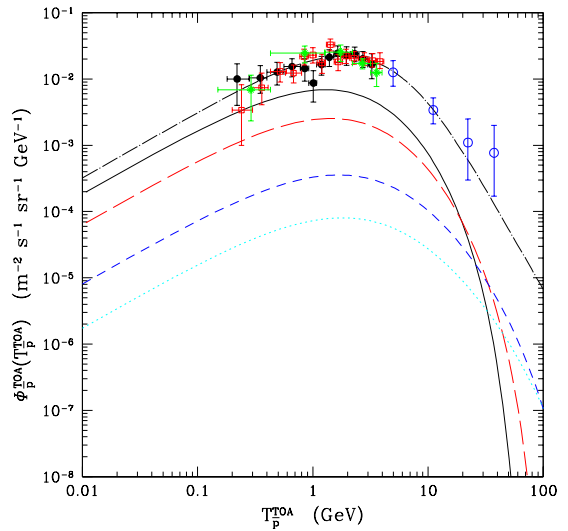}
\vspace{-37pt}
\caption{Primary TOA antiproton fluxes as a function of the antiproton
kinetic energy. The dashed line denotes the secon\-dary component, due
to spallation of cosmic rays ({\em i.e.} the background) taken from
Refs. \cite{PaperII,revue}. The other lines are representative fluxes
from neutralino annihilation of different masses \cite{pb}: the
solid line refers to $m_\chi = 60$ GeV, the long--dashed line to
$m_\chi = 100$ GeV, the short--dashed line to $m_\chi = 300$ GeV and
the dotted line to $m_\chi = 500$ GeV. The astrophysical parameters
for galactic propagation assume the best--fit values according to the
analysis of Refs. \cite{PaperI,pb}. Solar modulation is calculated
for a period of minimal solar activity. Full circles show the {\sc
bess} 1995--97 data \cite{bess95-97}; the open squares show the {\sc
bess} 1998 data \cite{bess98}; the stars show the {\sc ams} data
\cite{ams98} and the empty circles show the {\sc caprice} data
\cite{caprice}.}
\label{fig:prim_sec_data_med_solmin}
\end{figure}

\begin{figure}[t] \centering
\vspace{-20pt}
\includegraphics[width=1.0\columnwidth]{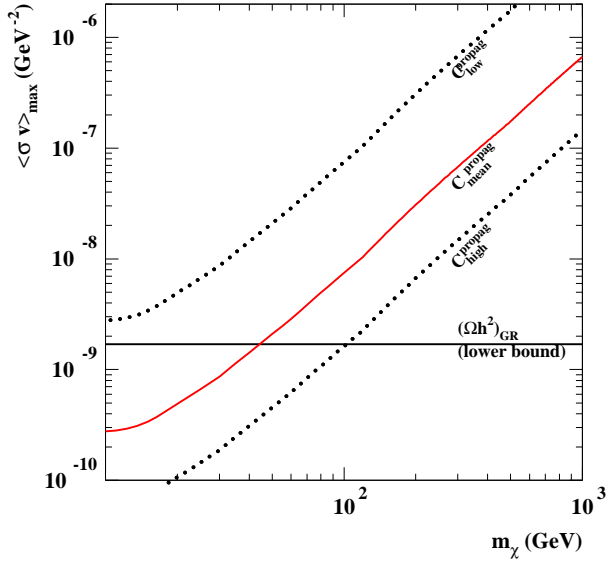}
\vspace{-20pt}
  \caption{The solid line shows the maximum value (upper bound) of
  $\langle\sigma_{\textrm{ann}}v\rangle_0$ that can be allowed by the
  cosmic antiproton data in the fist energy--bin of {\sc BESS} and
  {\sc AMS} ($T^{\rm TOA}_{\rm p} = 0.23$ GeV.)
  $\langle\sigma_{\textrm{ann}}v\rangle_0$ is the thermal average in
  the galactic halo of the WIMP annihilation cross section times
  relative velocity. The upper bound is shown as a function of the
  WIMP mass $m_\chi$ and has been derived using a median set of
  astrophysical parameters for the antiproton propagation. The upper
  and lower dotted lines show the uncertainty in the determination of
  the upper limit on $\sigmavzero$ coming from the astrophysical
  uncertainties. The horizontal solid line shows the lower bound on
  $\sigmavzero$ coming from the relic density of cold dark
  matter in standard cosmology, under the hy\-po\-the\-sis that the WIMP 
  annihilation cross--section
  in the early Universe $\sigmav$ equals $\sigmavzero$ (like in the
  case of $s$--wave dominance).}
\label{fig:sigpblim}
\end{figure}

The antiproton component of cosmic rays has been measured by many
detectors in space and the most recent results come from {\sc BESS},
{\sc AMS} and {\sc CAPRICE} ex\-pe\-ri\-ments. The standard production of
cosmic antiprotons from spallation of nuclei on the diffuse Milky--Way
gas is enough to explain the data \cite{PaperII,revue}, but the
error--bars leave a small room also for an exotic antiproton
signal. Such a signal could come from the annihilation of WIMPs in the
Galaxy (see {\em e.g.}  Ref. \cite{pb}). This situation is shown in
Fig. \ref{fig:prim_sec_data_med_solmin}, where the experimental data
are plotted together with the background component and some examples
of a signal coming from WIMP annihilation, as calculated in
Ref. \cite{pb}.

The antiproton flux produced in a point of cylindrical coordinates
$(r,z)$ in the galactic halo, where the dark matter density is
$\rho(r,z)$, depends on the annihilation cross section of the WIMPs
averaged over their ve\-lo\-ci\-ty distribution in the galactic halo
$\sigmavzero$ and on the final--state branching--ratios into the different
possible final sates $F$ \cite{pb}:
\begin{equation}
q_{\bar p}(r,z:T_{\bar p}) = \sigmavzero \frac{\rho^2(r,z)}{2
m_\chi^2}\, \sum_F {\rm BR}(\chi\chi \rightarrow F) \,
\left(\frac{dN^F_{\bar p}}{dT_{\bar p}}\right)
\label{eq:pbar}
\end{equation}
where $T_{\bar p}$ is the antiproton kinetic energy. The antiprotons
then propagate and diffuse in the galactic medium until they reach the
Earth position in the Galaxy of coordinates $(R_\odot,0)$:
\begin{equation}
q_{\bar p}(r,z:T_{\bar p})\, \longrightarrow \, \Phi_{\bar
p}(R_\odot,0:T_{\bar p})
\label{eq:pbar1}
\end{equation}
The process of propagation has been discussed in details in
Ref. \cite{pb}, where it has been shown that, contrary to the case of
the spallation antiprotons \cite{PaperI}, a large uncertainty in the
low--energy tail of the antiproton signal is present, due to the current
uncertainties on the astrophysical parameters which describe the
diffusion and propagation processes \cite{pb,PaperI}. This uncertainty
is about one order of magnitude up or down around the best fit result
\cite{pb}, for antiproton kinetic energies below 1 GeV where a signal
may be more promisingly approached for WIMPs in the tens--hundreds of
GeV mass range, and without the need of boosted enhancements on the
dark matter density \cite{pb,boost}. The additional solar modulation effect
introduces another source of uncertainty which is expected to be 
less important than
the one due to galactic diffusion. Also the uncertainty coming from
the dark matter profile is not very relevant, since antiprotons which
reach the Earth are produced relatively close in the Galaxy, due to
diffusion, and therefore they do not strongly feel the quite uncertain
galactic--center mass distribution. Differences in the galactic halo
shape affect the antiproton signal at most 20\% \cite{pb}. For a
thorough discussion of all these topics and of the calculation of the
antiproton signal which we also use in the present analysis, we refer
to Ref. \cite{pb}.

The antiproton signal is a powerful tool for constrai\-ning WIMP dark
matter, since the amount of antiprotons produced by typical WIMPs
which can account for the dark matter abundance of Eq. (\ref{oh2
constraint}) is at the level of the background and may even exceed it
sizably \cite{pb,pbarlim}. The astrophysical uncertainties limit
somehow the capabilities of this type of signal, but nevertheless
combined limits in the astrophysical--particle physics parameter space
may be set, especially for relatively light WIMPs. Ref. \cite{pbarlim}
showed how WIMPs in the mass range from few GeV to hundreds of GeV
are currently constrained by antiprotons searches. The possibility to
set bounds mainly relies on the fact that the low--energy tail of the
predicted antiproton signal may exceed the room left in the 
experimental uncertainty of the data over the calculated background. 
We make use
of the same argument in the present discussion to use the antiproton
signal to set limits on $\sigmavzero$ and to transform these limits on
bounds on the enhancement of the Hubble rate around WIMP freeze--out
under the condition that the WIMPs satisfy the cosmological bound on
the amount of cold dark matter.

A few comments are in order here before we proceed to discuss our
strategy. First of all, we explain why we do not use also other
indirect detection signals other than antiprotons. WIMP annihilation
may obviously also produce gamma rays and positrons. However, contrary
to the case of antiprotons which, almost independently of the halo
density profile, are naturally at the level of the background and
experimental data for cosmologically relevant WIMPs, both positrons
and gamma rays usually require sizeable boosts in the dark matter
density in order to reach detectable levels. This introduces a strong
model--dependent variable which does not allow us to use these indirect
signals as reliable tools for setting limits. For instance, a limit
obtained from gamma--rays would fade out completely unless a very
steep density profile is present at the galactic center. The same
occurs for positrons, which need strongly clumped structures very
close to our position in the Galaxy. Even though these indirect
signals are very appealing for dark matter stu\-dies, they do not prove
to be useful in the analysis we want to carry on in this paper. Very
promising will potentially be antideuterons \cite{dbar}, but for that
we have to wait for the foreseen experimental set--ups able to access 
the required sensitivities \cite{gaps,ams}.

For similar reasons, we are concentrating our ana\-ly\-sis on the
low--energy tail of the antiproton flux: also at energies in the tens
of GeV range (where {\sc CAPRICE} data are available) a signal could
manifest itself above the background, which is here fast
decaying. Moreover, astrophysical uncertainties are less
relevant at these ener\-gies. However, in order to have large signals in 
this range of
ener\-gies, we need a suitable WIMP number density, which can be likely
obtained only with some degree of over--density (due to the fact that we
need here heavier WIMPs, whose number density is damped by the
$m^{-2}$ factor) \cite{pb,sweden}. Again, since we do not want to add
additional arbitrary inputs in our analysis (the boost factor), we
focus on the low--energy tail.

Finally, we must remind that the antiproton signal depends on (and
therefore can be constrained to set bounds on) $\sigmavzero$, while
the relic abundance depends on $\sigmav$, {\em i.e.} on the thermal
average on the annihilation cross section at a different (larger)
temperature. A constraint on $\sigmavzero$ is not therefore directly
transferable to a bound on the WIMP relic abundance, and
vice--versa. The usual expansion of the thermal average:
\begin{equation}
\langle\sigma_{\textrm{ann}}v\rangle = a + \frac{b}{x} + \cdot\cdot\cdot
\label{temp expansion sigma}
\end{equation}
holds in most of the cases, noticeable differences are when
co--annihilation effects are present \cite{coannihil,pole1} or when
annihilation occurs close to a resonance or a threshold \cite{pole2}.
However, in most of the typical cases, the thermal average of the annihilation
cross section is mildly temperature--dependent. We will first discuss
through the paper the case of temperature--independence, {\em i.e.}:
$\sigmav=\sigmavzero=a$. We will then discuss how our results are
changed when the first--order expansion of Eq. (\ref{temp expansion
sigma}) is relevant. The most important effect on our results
basically comes from the difference between the freeze--out and
current value of $\sigmav$, {\em i.e.} on the ratio:
\begin{equation}
{\cal R} = \frac{\sigmav_{T_{\rm f}}}{\sigmavzero}
\label{eq:ratio}
\end{equation}
Therefore, an estimate of how our results would change, for instance
when the relic abundance is determined by co--annihilation effects, is
to assume a given value for ${\cal R}$ in relating $\sigmav$ at freeze--out and
$\sigmavzero$. We also need to comment that, in order to apply the antiproton 
bound, the WIMP annihilation must proceed sizably to non-leptonic final 
states. In fact, for annihilation into leptons, antiprotons are not produced 
and our bounds are loosened by a factor given by the branching ratio into 
non-leptonic final states.

Let us now determine the upper limit on the WIMP annihilation cross
section derived from the observational upper limit on the exotic
cosmic antiproton flux. We do not attempt here a statistical analysis
of the data as was done in Ref. \cite{pbarlim}. We instead use a
simpler approach which uses the most relevant information coming from
the antiproton data and at the same time allows us to have an insight
on the results, which could instead be more difficult to obtain
by a statistical treatment of the data. This approach was the one also
adopted in Ref. \cite{pb}. We consider the experimental result in the
lowest--energy bin at $T_{\bar{p}}^{\textrm{TOA}} = 0.23$ GeV. By
subtracting the background \cite{PaperII}, we are left with a 90\%
C.L. upper limit for the exotic antiproton component at the top of the
atmosphere (TOA) of: $\Phi_{\bar{p}}^{{\textrm{TOA}}} = 2\cdot10^{-3}
\textrm{m}^{-2} \textrm{s}^{-1} \textrm{sr}^{-1} \textrm{GeV}^{-1}$
\cite{pb}. In order to find the corresponding upper bound on the WIMP
annihilation cross section we must specify our assumptions in the
calculation of Eq. (\ref{eq:pbar},\ref{eq:pbar1}) of the expected antiproton
flux for any given mass $m_\chi$ and cross section $\sigmavzero$.

We assume an isothermal halo model with core radius 3.5 kpc and local
dark matter density 0.3 GeV $\textrm{cm}^{-3}$. When nothing else is
mentioned we use the mean values for the set of astrophysical
parameters of the propagation and diffusion models given in
Ref.\cite{pb}, but we discuss also the set of parameters which provide
the maximal and minimal antiproton flux, in order to properly take
into account this intrinsic and dominant source of uncertainty. As
for the antiproton spectrum, we assume for definiteness that the WIMP
annihilate only into a $\bar{b}b$ pair, and we therefore use the
corresponding spectra given in Ref. \cite{pb}. A more generic
annihilation final state would only mildly change our results (except for 
the already--mentioned leptonic final state).

The result on the upper bound on $\sigmavzero$ is shown in
Fig.~\ref{fig:sigpblim} as a function of the WIMP mass and for the
median, as well as maximal and minimal set of astrophysical
parameters. We notice that the astrophysical uncertainty is severe,
but nevertheless allows us to set limits on the maximal amount of
$\sigmavzero$ which is allowed in order not to go in conflict with the
experimental data. The horizontal solid line instead shows the lower
limit on $\sigmav$ coming from the cosmological bound on the WIMP relic
abundance of Eq. (\ref{oh2 constraint}), in the case of temperature
independent thermal average and standard cosmology. The figure clearly
shows that there is tension between the cosmological bound and the
antiproton limit: in the case of the median bound from antiproton, we
see that WIMPs lighter than about 50--60 GeV lead to an antiproton
signal which is in excess of the experimental bound. This tension is
released if we consider the minimal set of astrophysical parameters
(upper dotted curve), but the figure clearly shows that antiprotons
are a powerful tool for setting limits. Ref. \cite{pbarlim} already
discussed in details this tensions and the corresponding limits. We
instead here use this argument to set bounds on cosmological models:
since the cosmological models we are discussing predict larger
lower--bounds on $\sigmav$, we see that the tension with antiproton
data is enhanced and limits can be set on the maximal
amount of enhancement of the Hubble rate at the time of WIMP
freeze--out.

\section{Constraining the Hubble rate with Dark Matter}
\label{sec:constraints}

In this Section we show how the combined constraints on the WIMP relic
density and antiproton signal can strongly constrain the Hubble rate
in the early Universe. Let us first recall our assumptions about the
Hubble rate and the WIMPs. We assume that the Hubble rate is enhanced
by a temperature--dependent factor $A(T)$ in the early Universe, for
which we consider the parametrization of Eq. (\ref{A def}). At the
temperature $T_\textrm{f}$ at which the WIMP would freeze out in the
standard case, the enhancement factor is equal to the parameter
$\eta$, $A(T_\textrm{f}) = \eta$ (for $\eta \gg 1$). The slope of
$A(T)$ is set by the parameter $\nu$. At a temperature $T =
T_\textrm{re}$, the expansion rate "re--enters'' the standard one. We
consider a general WIMP model cha\-rac\-te\-rized solely by a mass,
$m_\chi$, and an annihilation cross section. We assume first that
$\langle\sigma_{\textrm{ann}}v\rangle=\sigmavzero$ is temperature
independent, and therefore it determines both the relic abundance and
the antiproton signal directly. The WIMP should explain the
observational amount of cold dark matter of Eq. (\ref{oh2
constraint}), {\em i.e.} it must be the dominant CDM component.

In the following we combine the relic density and antiproton
constraints on the WIMP to obtain the upper bound on the enhancement
of the Hubble rate. As we have discussed, this is possible because the
Hubble rate together with the WIMP mass and cross section determines
the WIMP abundance, and at the same time the WIMP cross section is
bounded from above by the cosmic antiproton data. In Section
\ref{subsec:eta} we find the mass dependent upper bound on $\eta$,
{\em i.e.} the enhancement factor at the time of the WIMP freeze
out. In Section \ref{subsec:relic ratio} we determine the
corresponding upper bound on the enhancement of the relic WIMP density
as compared to standard cosmology. Finally, in Section
\ref{subsec:relax assumptions} we analyze what happens to the bound on
$\eta$ when some of our assumptions about the Hubble rate and the WIMP
model are changed.

\subsection{Constraining the expansion rate}\label{subsec:eta}

Let us in this section show our numerical results on the Hubble rate 
constraints derived from the relic density and antiproton bound. Our 
parametrization of the Hubble
enhancement function Eq. (\ref{A def}) has three free parameters:
$\eta$, $\nu$ and $T_\textrm{re}$ (while $T_\textrm{f}$ is set by the
WIMP mass and cross section through Eq. (\ref{eq:xf})). As we shall
see at the end of this section, the enhancement is basically
independent of $T_\textrm{re}$, as long as $T_\textrm{re}\ll T_{\rm
f}$. This can be understood also by means of the analytical
approximations given in Appendix \ref{app:analytic}. We choose to set
$T_{\textrm{re}}=10^{-3}$ GeV throughout the paper as a reference
value. This corresponds to the lowest value which can be safely assumed
to be compatible with big bang nucleosynthesis
\cite{scalartensorrelic}. The enhancement of the Hubble rate is then
studied in the two-dimensional parameter space $(\eta,\nu)$.

Let us now construct an exclusion plot in the $(\eta,\nu)$ plane. We
first fix the WIMP mass at $m_\chi = 500$ GeV as a reference
value. Then, for each point in the $(\eta,\nu)$ plane we make a scan
in the WIMP annihilation cross section
$\langle\sigma_{\textrm{ann}}v\rangle$. The cross section together
with the mass fixes $T_\textrm{f}$, which is the temperature at which
the WIMP would freeze out in the standard cosmology scenario. We then
solve the Boltzmann equation with the enhanced Hubble rate. We can now
apply the first bound, namely the constraint Eq. (\ref{oh2
constraint}) on the relic dark matter density. This constraint selects
a small interval in $\langle\sigma_{\textrm{ann}}v\rangle$ for each
$(\eta,\nu)$. Finally we apply the constraint coming from the cosmic
antiproton data on $\sigmavzero$, by means of the result shown in
Fig.\ref{fig:sigpblim}, which gives us the upper bound on the
annihilation cross section for any given mass.

%%% --------------------------------------------------------------------------------
%
\begin{figure}[t] \centering
\vspace{-20pt}
\includegraphics[width=1.0\columnwidth]{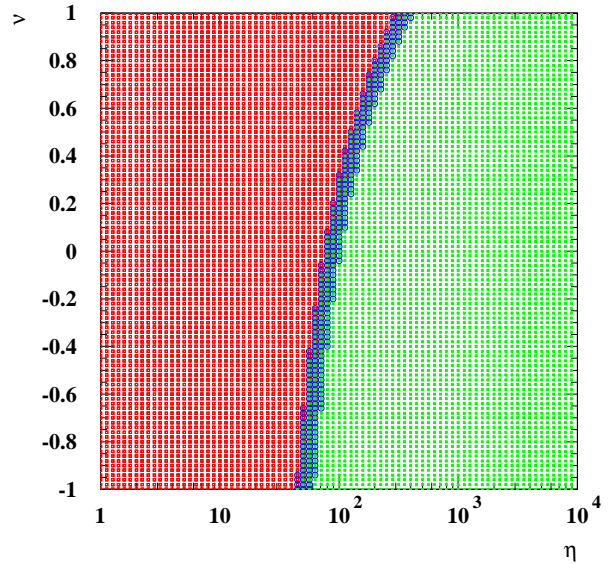}
\vspace{-20pt}
  \caption{Exclusion plot in the $(\eta,\nu)$ plane for a WIMP mass of
  $m_\chi = 500$ GeV. $\eta$ and $\nu$ are two of the parameters of
  the Hubble enhancement function. $\eta$ is the enhancement factor at
  the freeze--out temperature of the WIMP. The ``re--entering''
  temperature has been fixed at $T_\textrm{re} = 10^{-3}$ GeV. The
  WIMP annihilation cross section $\sigmav$ is chosen for each point
  in order to fulfill the relic density constraint. This cross section
  has then been compared to the antiproton upper bound on the cross
  section. The central almost vertical band denotes points which have
  exactly the limiting value of the cross section; the points on the
  left of the band refer to cross sections below the antiproton upper
  bound; the points on the right of the band refer to cross section in
  excess of the antiproton upper bound.}
\label{fig:enplane06}
\end{figure}

In fig.~\ref{fig:enplane06} we show the exclusion plot in the
$(\eta,\nu)$ plane. The points in the central almost--vertical band
refer to annihilation cross sections
$\langle\sigma_{\textrm{ann}}v\rangle$ equal to the maximum value
allowed by the antiproton data (using the median set of astrophysical 
parameters) as shown in
Fig.~\ref{fig:sigpblim}. The points on the left of the vertical band
all refer to smaller values of the cross section and thus
sa\-tis\-fy the antiproton bound. On the contrary, the points on the right
are in disagreement with the antiproton data as they require a high
cross section in order to fulfill the density constraint. Notice that
the cross section is not uniquely determined by the density
constraint, since for fixed values of $\eta$, $\nu$ and $m_\chi$ we
can slightly vary the annihilation cross section to move around in the
allowed density interval of Eq. (\ref{oh2 constraint}). Similarly,
points next to each other can have identical cross section and fulfill
the density constraints with just slightly different values of the
relic abundance \footnote{Note that models that have the same WIMP
mass and cross section would have identical relic abundance in
standard cosmology. Here, however, we are interested in the relic
abundance in a cosmology with enhanced Hubble rate, and this abundance
does depend also on the parameters $\eta$ and $\nu$.}. Thus, the width
of the vertical band, that marks the antiproton bound, is due to the
allowed interval of the relic abundance. Also, in the vertical band we
do not only have models with a cross section equal to the antiproton
limit, but also models with a slightly lower or higher cross section.

From fig.~\ref{fig:enplane06} we conclude that the constraint coming
from the cosmic antiprotons produced by dark matter annihilation can
be used to put constraints on the Hubble rate in the very early
Universe. The figure shows that, for $m_\chi = 500$ GeV, we get an
upper bound on the enhancement factor $\eta$ of the Hubble rate at the
WIMP freeze--out of around $10^2$ (with some dependence on the actual
values of the slope parameter $\nu$). This bound on $\eta$ may be
considered as significant: in Ref. \cite{scalartensorrelic} we found
perfectly viable scalar--tensor models with enhancement factors of up
to $10^4$ at the WIMP freeze--out temperature, once the BBN, CMB and
gravitational probe limits were satisfied. Our results shows that
alternative, and even stronger, bounds on dark energy models can
therefore be obtained also by looking at the imprints left on the dark
matter properties by these modified cosmologies at the time dark
matter formed in the early Universe, if we assume that dark matter is
provided by a thermal relic.

%%% --------------------------------------------------------------------------------
%
\begin{figure}[t] \centering
\vspace{-20pt}
\includegraphics[width=1.0\columnwidth]{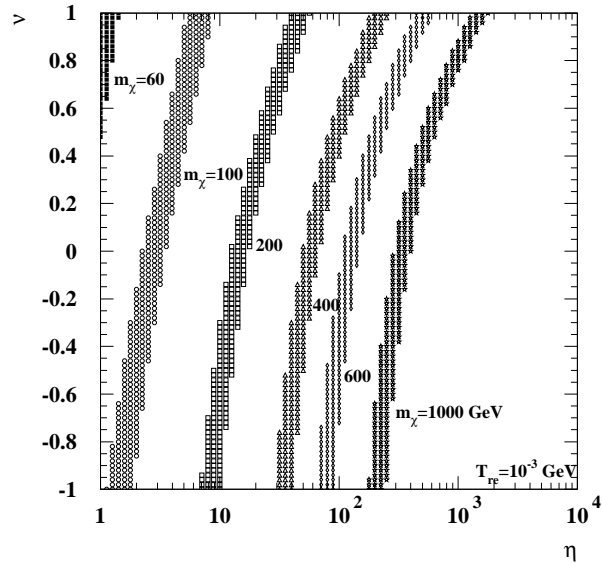}
\vspace{-20pt}
  \caption{The same as Fig.~\ref{fig:enplane06} but including the
  result for dif\-fe\-rent WIMP masses. Going from left to right the
  vertical bands correspond to the antiproton bound for WIMP masses of
  60 GeV, 100 GeV, 200 GeV, 400 GeV, 600 GeV and 1000 GeV
  respectively. For any given mass, the vertical band marks the limit
  between the excluded part of the parameter space to the right and
  the part to the left which is in agreement with the antiproton
  data. The ``re--entering'' temperature is $T_\textrm{re} = 10^{-3}$
  GeV.  }
\label{fig:enmass07}
\end{figure}

The results in fig.~\ref{fig:enplane06} were calculated for a fixed
WIMP mass of 500 GeV. Let us now explore how the bound on $\eta$
depends on the WIMP mass. The result is shown in
fig.~\ref{fig:enmass07}. Each vertical band corresponds to the maximum
allowed cross section for any given mass according to
fig.~\ref{fig:sigpblim}. Going from left to right in
fig.~\ref{fig:enmass07}, the vertical bands corresponds to the masses
60 GeV, 100 GeV, 200 GeV, 400 GeV, 600 GeV and 1000 GeV
respectively. For any given mass, the part of the $(\eta,\nu)$ plane
to the right of the vertical band is excluded by the antiproton data
while the part to the left is allowed. The constraint on $\eta$ is
very strong, especially for low WIMP masses. In particular, the figure
shows that WIMPs lighter than about 60 GeV are close to being
excluded.  A word of caution is in order here: to make a precise claim
on a lower bound on the WIMP mass would require a more thorough
analysis, since we should include the uncertainty on the propagation
and diffusion parameters for the calculation of the antiproton flux
and also apply a more refined statistical analysis on the antiproton
data. We will show the effect coming from the astrophysical parameters
in Sect. \ref{subsec:relax assumptions}, and we anticipate that for
the less stringent set of parameters, WIMPs lighter than 60 GeV are
perfectly viable. These results are in agreement with what is found in
Ref. \cite{pbarlim} where a careful analysis of the bounds that can be
set to the WIMP parameters by antiproton data in standard cosmology
has been performed.

%%% --------------------------------------------------------------------------------
%
\begin{figure}[t] \centering
\vspace{-20pt}
  \includegraphics[width=1.0\columnwidth]{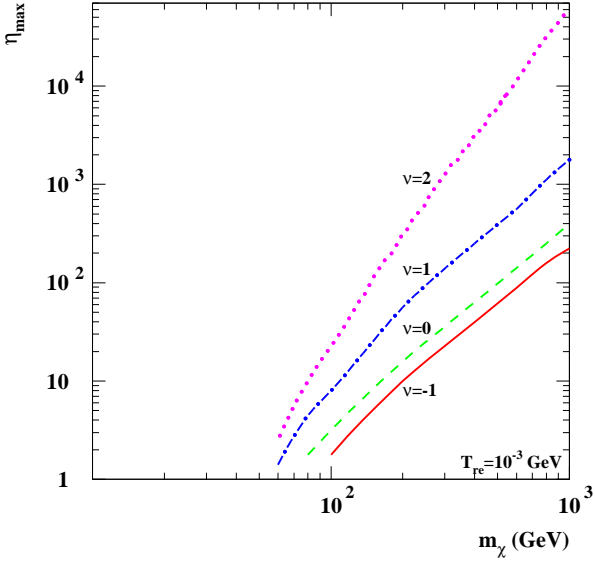}
\vspace{-20pt}
  \caption{Upper limit on the value of $\eta$ obtained by imposing the
  combined constraints on the WIMP relic abundance and the WIMP
  antiproton signal. $\eta$ is the Hubble enhancement factor at the
  time where the WIMP freezes out in standard cosmology. The bound on
  $\eta$ is shown as a function of the WIMP mass for the slope
  parameter of the enhancement function $\nu = -1, 0, 1, 2$. The
  parameter $T_\textrm{re}$ has been set to $10^{-3}$ GeV.
  $T_\textrm{f}$ is determined by the WIMP mass and cross section by
  means of the freeze--out condition of Eq. (\ref{eq:xf}). The WIMP
  cross section is fixed at the limit of Fig.~\ref{fig:sigpblim}.}
\label{fig:etamax03large}
\end{figure}
\begin{figure}[t] \centering
\vspace{-20pt}
\includegraphics[width=1.0\columnwidth]{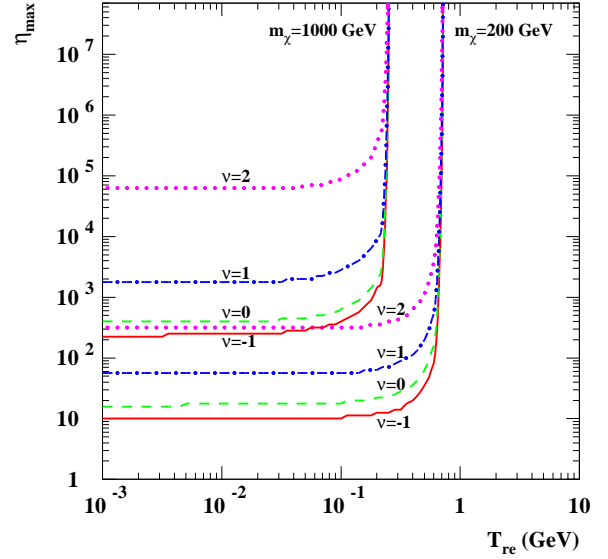}
\vspace{-20pt}
  \caption{Upper bound on the Hubble enhancement parameter $\eta$ as a
  function of $T_\textrm{re}$, {\em i.e.} the temperature at which the
  Hubble expansion ``re--enters'' the standard evolution. The upper
  bound on $\eta$ is derived from the combination of the constraint on
  the WIMP relic density and antiproton signal. The numerical
  derivation is done for two different WIMP masses and four different
  values of the parameter $\nu$. The masses are $m_\chi = 200$ GeV and
  $m_\chi = 1$ TeV corresponding to $T_\textrm{f} \sim 7.46$ GeV
  respectively $T_\textrm{f} \sim 31.86$ GeV for the limiting cross
  section. The upper set of curves is for the high mass. For each of
  the two set of curves we have $\nu = 2,1,0,-1$ when going from the
  top to the bottom.}
\label{fig:trevetamax_c}
\end{figure}

From fig.~\ref{fig:enmass07} we see that the upper bound on $\eta$
increases less than an order of magnitude when the parameter $\nu$ is
varied from $-1$ to $+1$ for any given mass. For large masses, the
increase can be two orders of magnitude when we continue to
$\nu = 2$. This can be seen in fig.~\ref{fig:etamax03large}, which
displays the mass dependence of the upper bound on $\eta$, as derived
for four different values of $\nu$.

The behaviour of the upper bound on $\eta$ can be understood also by
means of the analytic approximations of Appendix \ref{app:analytic}.
The conditions we want to satisfy at the same time are the following:
\begin{eqnarray}
\Omega h^2 &\leq& (\Omega h^2)_{CDM} \quad\, \mbox{cosmological bound} \\
\sigmavzero &\leq& \sigmavzero^{\bar p}\quad\quad\, \mbox{antiproton bound}
\end{eqnarray}
where $(\Omega h^2)_{CDM}$ is the maximal allowed value of the relic
abundance constraint of Eq.~(\ref{oh2 constraint}) and
$\sigmavzero^{\bar p}$ is the upper limit on the annihilation cross
section coming from Fig.~\ref{fig:sigpblim}. This limit may be
approximated as: $\sigmavzero^{\bar p} \simeq 0.95 \cdot
10^{-12}\,m^{1.95}_{\chi}$ GeV$^{-2}$ for $m_\chi \gsim 60$ GeV. The two bounds
transform in the following condition:
\begin{equation}
f(\eta) \leq \sigmavzero^{\bar p} \, (\Omega h^2)_{CDM}
\label{eq:feta bound} 
\end{equation}
where, by means of the approximations given in Appendix \ref{app:analytic}:
\begin{equation}
f(\eta) = {\cal C} \left [
            {\cal I}(x_{\rm f},\hat x,\eta,\nu) + r_G\, \frac{1}{\hat x} 
        \right ]^{-1}
\label{eq:feta}
\end{equation}
and ${\cal C} = s_0\,\rho_c^{-1}\,B^{-1}\,{\cal G}^{-1}(x_f)$. The maximal
value of $\eta$ occurs when $\eta$ saturates the bound of
Eq. (\ref{eq:feta bound}). It is easy to show that:
\begin{eqnarray}
\eta_{\rm max} &=& 
 \left[\frac{\sigmavzero^{\bar p}\,(\Omega h^2)_{CDM}}{{\cal C} (\nu-1)\, x_{\rm f}} \right]^{\nu} 
 \sim m^{1.95\cdot\nu}_\chi \, \mbox{for $\nu>1$} \nonumber \\
\eta_{\rm max} &=& 
 \left[\frac{\sigmavzero^{\bar p}\,(\Omega h^2)_{CDM}}{{\cal C} (1-\nu)\, x_{\rm f}} \right]
 \sim m^{1.95}_\chi \, \mbox{for $\nu<1$}
\label{eq:etamax}
\end{eqnarray}
and a slightly more complicated expression for $\nu=1$ (kination case).
These behaviours are confirmed by our numerical calculations, where we
approximately find $\eta_{\rm max}\propto m_\chi^{2.1}$ for $\nu =
-1,0,1$ (with a slight dependence on which masses are used for the
calculation) and $\eta_{\rm max}\propto m_\chi^{3.5}$ for $\nu =
2$. 

We have shown in this Section that we can use the combined constraints
on the WIMP relic density and antiproton production to put an upper
limit on $\eta$, {\em i.e.} we can constrain the Hubble expansion at a
very early time in the history of the Universe, as $\eta =
A(T_\textrm{f})$. The exact temperature at which we constrain the
expansion in our discussion, {\em i.e.} the standard freeze out
temperature $T_\textrm{f}$, does however depend on the WIMP
mass. Approximately, $x_\textrm{f}$ is always of the order of 25
($x_\textrm{f} = m_\chi/T_\textrm{f}$). Some numerical examples are:
$T_\textrm{f}(m_\chi=60\,\textrm{GeV}) \simeq 2.6$ GeV,
$T_\textrm{f}(m_\chi=200\,\textrm{GeV}) \simeq 7.5$ GeV,
$T_\textrm{f}(m_\chi=600\,\textrm{GeV}) \simeq 20$ GeV and
$T_\textrm{f}(m_\chi=1000\,\textrm{GeV}) \simeq 32$ GeV, where we have
used the upper antiproton bound for the value of the WIMP annihilation
cross section.

Fig.~\ref{fig:trevetamax_c} shows the upper bound on $\eta$ as a
function of $T_\textrm{re}$ for two different values of the WIMP mass
and four different values of the parameter $\nu$. Models that stay
below the upper bound of $\eta$ would give WIMPs that either are
under-abundant or produce less antiprotons than the maximally allowed
flux. Fig.~\ref{fig:trevetamax_c} shows that the upper bound on
$\eta$ is independent of $T_\textrm{re}$ as long as the latter is much
smaller than the freeze out temperature. When the ``re--entering''
temperature is around one or two order of magnitudes below the
standard freeze out temperature, $\eta_{\rm max}$ starts to increase
and then becomes practically unbounded from above. In other words, one
can increase the Hubble rate arbitrarily if this is done for a short
amount of time. We see from fig.~\ref{fig:trevetamax_c} that the
``re--entering'' temperature at which $\eta$ starts to get unbounded
is lower for the higher mass case despite the fact that the freeze out
temperature is higher for $m_\chi = 1000$ GeV than for $m_\chi = 200$
GeV. This can be understood from the analytic solution of the modified
Boltzmann equation. We compare the contribution from the integration
from the modified freeze out until the ``re--entering'' temperature
({\em i.e.} the time where we still have enhanced expansion) with the
contribution from the integration from the ``re-entering'' temperature
untill today. The first contribution dominates when the
``re--entering'' temperature is very low. The analytic analysis shows
that the two contributions becomes equal approximately when $(\hat
T_{\textrm{f}}-T_{\textrm{re}})/T_{\textrm{re}} = \eta
[h_{\textrm{eff}}(x_{\textrm{re}})\sqrt{g_{\textrm{eff}}(\hat
x_{\textrm{f}})}] /[h_{\textrm{eff}}(\hat
x_{\textrm{f}})\sqrt{g_{\textrm{eff}}(x_{\textrm{re}})}]$, where $\hat
T_{\textrm{f}}$ is the freeze out temperature in the modified scenario
and where we have set $\nu = 0$ for definiteness. If $\eta$ stays
constant then $\Omega{h^2} \into (\Omega{h^2})_{\textrm{std}}$ when
$(\hat T_{\textrm{f}}-T_{\textrm{re}})$ shrinks to
zero. Alternatively, to avoid the WIMP to become under--abundant we
have to increase $\eta$ indefinitely. Since $\hat T$
grows approximately linearly with mass ($\hat T_{\textrm{f}} =
m_\chi/{\hat x_{\textrm{f}}}$ and $\hat x_{\textrm{f}}$ depends
logarithmically on the mass and the cross section) while the upper
bound on $\eta$ (fig. \ref{fig:etamax03large}) is approximately
proportional to the mass squared, it follows that $\eta$ becomes
unbounded at lower $T_{\textrm{re}}$ for $m_\chi = 1000$ GeV compared
to $m_\chi = 200$ GeV.

%%% --------------------------------------------------------------------------------
\begin{figure}[t] \centering
\vspace{-20pt}
\includegraphics[width=1.0\columnwidth]{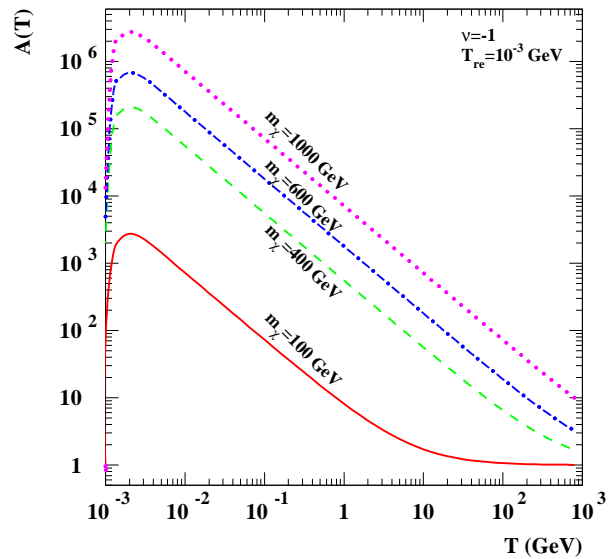}
\vspace{-20pt}
  \caption{Maximum allowed values of the enhancement function $A(T)$
  as a function of the temperature for our parametrization with $\nu =
  -1$ (close to a scalar--tensor cosmology). Notice that time is
  running from right to left. The limit is derived from $\eta_{\rm
  max}$. The solid red line refers to $m_\chi = 100 \ \textrm{GeV}$,
  the dashed green line to $m_\chi = 400 \ \textrm{GeV}$, the
  dash--dotted blue line to $m_\chi = 600 \ \textrm{GeV}$ and the
  dotted purple line to $m_\chi = 1\ \textrm{TeV}$. For all the
  curves we use $T_\textrm{re} = 10^{-3}$ GeV. The standard WIMP
  freeze--out temperature is given by the WIMP mass and the antiproton
  bound for the WIMP annihilation cross section.}
\label{fig:avt3}
\end{figure} 

\begin{figure}[t] \centering
\vspace{-20pt}
\includegraphics[width=1.0\columnwidth]{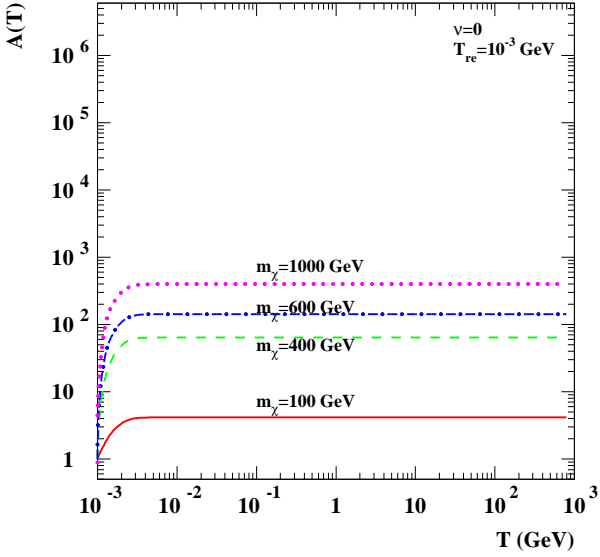}
\vspace{-20pt}
  \caption{Same as Figure \ref{fig:avt3} but for $\nu = 0$ (overall
  enhancement of Hubble rate, with standard temperature evolution).}
\label{fig:avt2}
\end{figure}

\begin{figure}[t] \centering
\vspace{-20pt}
\includegraphics[width=1.0\columnwidth]{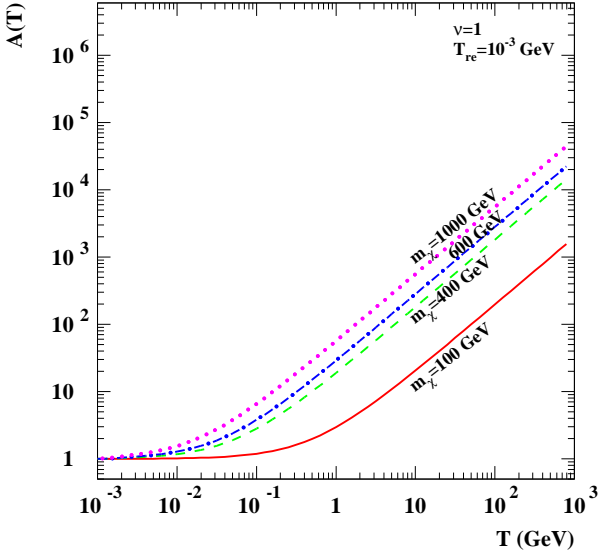}
\vspace{-20pt}
  \caption{Same as figure \ref{fig:avt3} but for $\nu = 1$ (kination case).}
\label{fig:avt1}
\end{figure}

\begin{figure}[t] \centering
\vspace{-20pt}
\includegraphics[width=1.0\columnwidth]{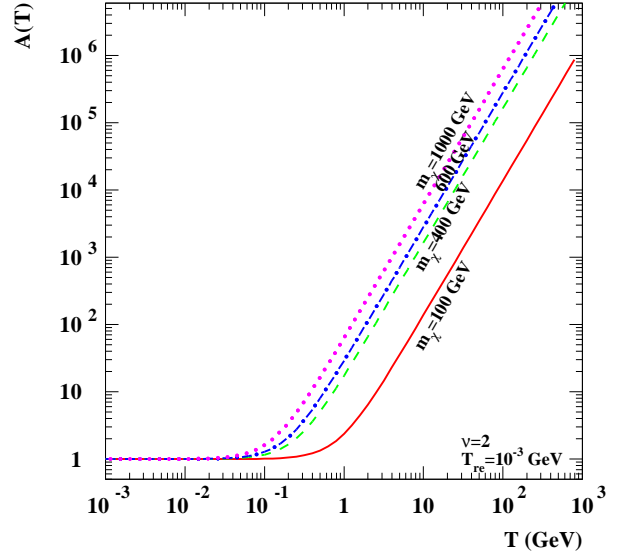}
\vspace{-20pt}
  \caption{Same as figure \ref{fig:avt3} but for $\nu = 2$ (RSII
  model \cite{Randall} of extra--dimension cosmology)}
\label{fig:avt4}
\end{figure}

The constraint on $\eta = A(T_\textrm{f})$ immediately constrains the
Hubble enhancement function $A(T)$ given in Eq.~(\ref{A def}) at any
temperature once we know the WIMP mass and the cosmological model
({\em i.e.} the parameters $\nu$ and $T_\textrm{re}$). In
Figs.~\ref{fig:avt3}, \ref{fig:avt2}, \ref{fig:avt1} and
\ref{fig:avt4} we show the maximum allowed enhancement function for
different values of the WIMP mass and the $\nu$ parameter. The curves
are derived for $T_\textrm{re} = 10^{-3} \ \textrm{GeV}$. As we just 
saw, the upper bound on $\eta$ does not depend on $T_\textrm{re}$ when 
$T_\textrm{re} \ll T_\textrm{f}$. Thus, a change of the re-entering 
temperature would solely change the temperature at which 
$A(T) \rightarrow 1$. 
Figure \ref{fig:avt3}, \ref{fig:avt2}, \ref{fig:avt1} and
\ref{fig:avt4} may be directly applied to constrain cosmological models 
with enhanced Hubble rate. For instance, for a scalar--tensor model like the
one studied in Ref. \cite{scalartensorrelic}, which refers to a
situation close to $\nu=-1$ (the actual value was $\nu=-0.82$) and
$T_{\rm re}=0.1$ GeV, by comparing Fig. \ref{fig:avt3} here and Fig. 6
of Ref. \cite{scalartensorrelic} we can conclude that the model of
Ref. \cite{scalartensorrelic} is compatible with antiproton data only
if the dark matter is composed by a WIMP with mass larger than 1 TeV.
This example shows how constraining the antiproton data may be on
otherwise viable dark energy models.

\subsection{The enhancement of the relic density}
\label{subsec:relic ratio}

In this section we derive upper limits on the enhancement of the WIMP
relic abundance in modified cosmologies as compared to the standard 
case, under the constraint that $\Omega h^2$ in the modified scenario
satisfies the bound of Eq. (\ref{oh2 constraint}). This constraint
alone is not actually enough to constrain the enhancement of the
density, as we shall see. When the antiproton bound for the WIMPs is
applied we find an upper bound on the density enhancement which is
practically independent of the parameter $\nu$, but that is of the
same order of magnitude as the upper bound on $\eta$.

As in the previous section we start by making a study in the
$(\eta,\nu)$ plane for a WIMP mass $m_\chi = 500$ GeV. As for
Fig. \ref{fig:enplane06}, we make a scan in
$\langle\sigma_{\textrm{ann}}v\rangle$ for each $(\eta,\nu)$ and keep
models where the WIMP density, as calculated with the enhanced Hubble
rate, fulfills the density constraint. For each of these models we
calculate also the WIMP density as it would be in the standard case
and we then find the ratio $R =
(\Omega{h}^2)/(\Omega{h}^2)_\textrm{GR}$.

%%% --------------------------------------------------------------------------------
%
\begin{figure}[t] \centering
\vspace{-20pt}
\includegraphics[width=1.0\columnwidth]{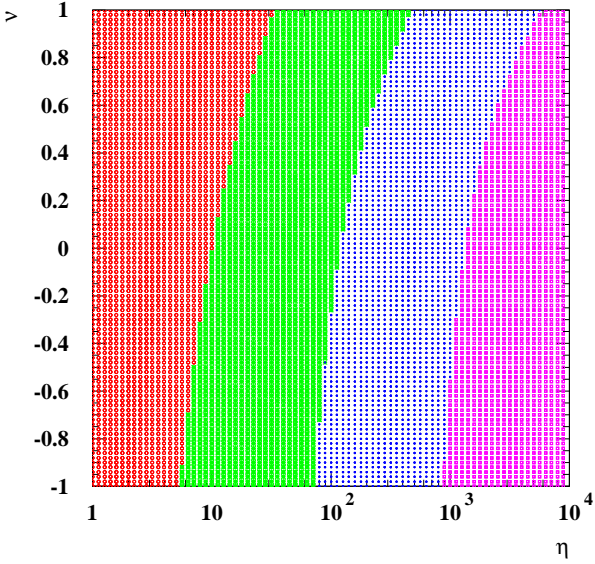}
\vspace{-20pt}
  \caption{Contour plot of the enhancement $R =
  (\Omega{h}^2)/(\Omega{h}^2)_\textrm{GR}$ of the WIMP relic abundance
  in a scenario with enhanced Hubble rate compared to the standard GR 
  cosmology. The different bands refer to (from left to right): $1\leq
  R \leq 10$, $10<R\leq 100$, $100<R\leq 1000$, $1000<R$. The highest
  value of $R$ is around $7.5\cdot{10}^3$. We have fixed $m_\chi =
  500$ GeV and $T_{\textrm{re}}=10^{-3}$ GeV. For all points, the WIMP
  relic--density, as calculated in the modified cosmology, satisfies
  the dark matter density constraint.}
\label{fig:evnoh2}
\end{figure}

Fig.~\ref{fig:evnoh2} shows contours of values of $R$. Going from low
to high values of $\eta$, the different regions correspond to $1\leq R
\leq 10$, $10<R\leq 100$, $100<R\leq 1000$, $1000<R$ with the highest
value being around $7.5\cdot{10}^3$. If we had continued to higher
values of $\eta$ we would have obtained even larger enhancements of
the WIMP density. From the previous section we know that $\eta$ can be
bounded from above by requiring that the WIMP fulfills the constraint
coming from the cosmic antiproton data. For $m_\chi = 500$ GeV the
bound on $\eta$ was found in Fig. \ref{fig:enplane06}. When we compare
this figure with the contours of $R$ in fig.~\ref{fig:evnoh2}, we
conclude that the antiproton data leads to the upper limit
$R\sim{100}$ for $m_\chi = 500$ GeV. We also see that the shape of the
contours of $R$ are very similar to the band in
Fig.~\ref{fig:enplane06} which marks the limiting cross section. We
therefore expect the upper bound on $R$ to be almost independent of
$\nu$. The final thing to note is that the numerical value of the
upper bound on $R$ and $\eta$ are of the same order of magnitude.

%%% --------------------------------------------------------------------------------
%
\begin{figure}[t] \centering
\vspace{-20pt}
\includegraphics[width=1.0\columnwidth]{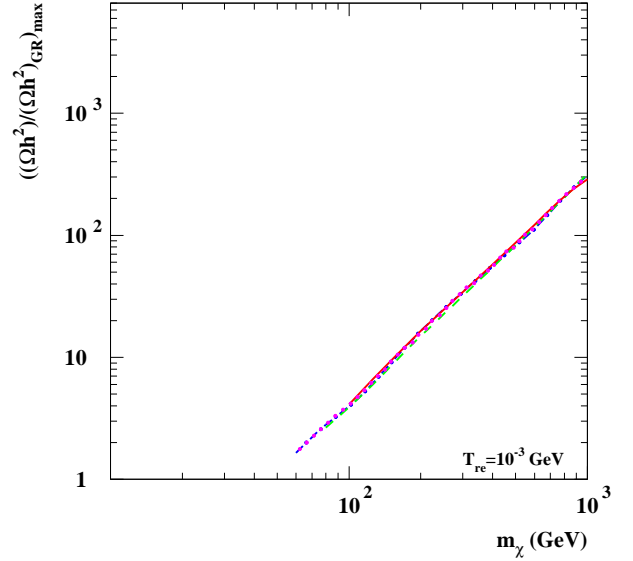}
\vspace{-20pt}
  \caption{Upper limit of $R =
  (\Omega{h}^2)/(\Omega{h}^2)_\textrm{GR}$ as a function of the WIMP
  mass. The bound comes from the combination of the WIMP relic density
  constraint and the antiproton bound. The lines, which practically
  fall on top of each-other, refer to: $\nu = -1$, $\nu = 0$, $\nu =
  1$ and $\nu=2$. The mass lower limit depend on $\nu$ (and it
  depends on the assumption of astrophysical parameters, which have
  been fixed to their median set in the figure). In the figure we have
  fixed $T_{\textrm{re}}=10^{-3}$ GeV.}
\label{fig:ratiomass_e}
\end{figure}

The upper bound on the relic density enhancement $R$ as a function of
the WIMP mass is shown in fig.~\ref{fig:ratiomass_e}, for the four
representative values of the parameter $\nu$. The approximate
behaviour $R\propto m_{\chi}^{1.9}$ can be understood by means of the
analytical approximation we already discussed.

\subsection{Changing the assumptions}
\label{subsec:relax assumptions}
\begin{figure}[t] \centering
\vspace{-20pt}
\includegraphics[width=1.0\columnwidth]{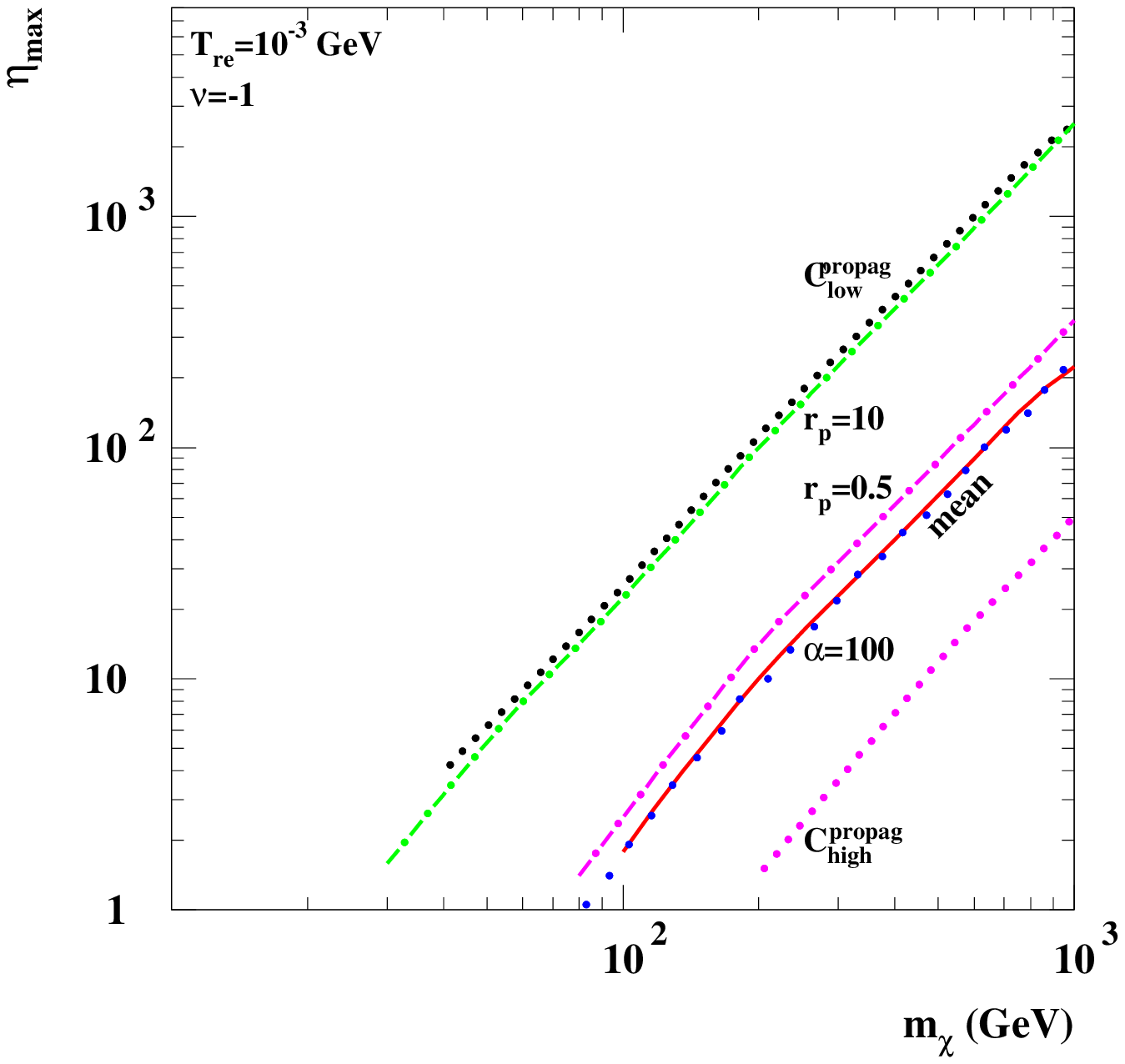}
\vspace{-20pt}
  \caption{Upper bound on $\eta$ as a function of the WIMP mass under
  different assumptions, as derived from the combined constraint on
  the WIMP relic abundance and antiproton flux. All curves have
  $T_{\textrm{re}} = 10^{-3}$ GeV and $\nu = -1$. The reference case
  is the red solid curve which is identical to our previous result in
  Fig.~\ref{fig:etamax03large}. The uppermost and lowermost dotted
  lines delimit the band due to the uncertainty on the astrophysical
  parameters for the antiproton propagation and diffusion. The
  uppermost curve is for the minimal set of astrophysical parameters
  while the lowermost is for the maximal set of parameters. The two
  dash--dotted lines, instead, include a $p$-wave contribution for the
  annihilation in the early Universe. For the upper green curve the
  $p$-wave factor is $r_p = 10$ (see text) and for the lower purple
  curve we have $r_p = 0.5$. The dotted line which falls on top of the
  red solid (mean) line is calculated with the Hubble enhancement
  function $A_2(T)$ with a change in the post-freeze--out temperature
  behaviour given by the parameter $\alpha = 100$.}
\label{fig:etamax04}
\end{figure}

In this Section we show how the upper bound on the Hubble enhancement
$\eta$ moves when some of our assumptions are changed. We are going
to consider three kinds of modifications: we shall consider a
temperature--dependent annihilation thermal cross section, by adding a
$p$-wave contribution; we change the parametrization of the Hubble
enhancement $A(T)$; finally we take into account the uncertainty of
the propagation and diffusion of the cosmic antiproton signal.

Let us first consider what happens if our assumption about pure
$s$-wave annihilation does not hold. In this case $\sigmav$, relevant
for the relic abundance, and $\sigmavzero$, relevant for the
antiproton calculation, are no longer equal. By adding a $p$-wave
term $b/x$ in the temperature expansion of Eq. (\ref{temp expansion
sigma}), $\langle\sigma_{\textrm{ann}}v\rangle$ turns out to be larger
(or smaller, depending on the sign of $b$) in the early Universe than
it is today where the $p$-wave contribution is negligible. The ensuing
limit we obtain is less (more) stringent than the one we would obtain
for pure $s$--wave annihilation, depending whether $b$ is positive or
negative. Fig.~\ref{fig:etamax04} shows the effect for two
representative cases in terms of $r_p=b/(a\,x_{\rm f})$: $r_p = 0.5$ 
(lower dash--dotted purple line) and $r_p = 10$ (upper dash--dotted 
green line). The last value is actually extreme and it has been considered 
to make an example. 

We notice that the results we obtain for a non negligible
$p$--wave case are similar to what we would obtain by assuming
a value for the ratio ${\cal R}$ in Eq. (\ref{eq:ratio}) of the order
of ${\cal R} = (1+\omega\,r_p)$, where $\omega\sim{\cal O}(1)$ depends
on details of the integrals given in Eq. (\ref{eq:sigmaint}), and
therefore also on the cosmological model (simple ana\-ly\-ti\-cal
expressions may be derived by means of the approxi\-mations of Appendix
\ref{app:analytic}). The cases shown in Fig. \ref{fig:etamax04} refer
to ${\cal R} \simeq 1.2-1.5$ for $r_p = 0.5$ and ${\cal R} \simeq 6-10$ for
$r_p = 10$. The figure therefore shows that the bound on $\eta_{\rm
max}$ is sensitive to large values of the ratio ${\cal R}$, as
expected. In the case, for instance, of coannihilations, for which
${\cal R}$ may be a large number (even some orders of magnitude) we
see that the bounds from antiprotons are much less stringent.
Approximately, the change in the upper bound on $\eta$ obtained in
Eqs.~(\ref{eq:etamax}) for the case of time--independent cross section, 
can now be rephrased as:
\begin{eqnarray*}
\eta_{\rm max} &=& 
 \left[\frac{{\cal R}\,\sigmavzero^{\bar p}\,(\Omega h^2)_{CDM}}{{\cal C} (\nu-1)\, x_{\rm f}} \right]^{\nu} 
 \quad \mbox{for $\nu>1$} \\
\eta_{\rm max} &=& 
 \left[\frac{{\cal R}\,\sigmavzero^{\bar p}\,(\Omega h^2)_{CDM}}{{\cal C} (1-\nu)\, x_{\rm f}} \right] 
 \quad\ \ \mbox{for $\nu<1$}
\label{eq:etamaxr}
\end{eqnarray*}
with, again, a slightly more complicated expression for $\nu=1$.

\begin{figure}[t] \centering
\vspace{-20pt}
\includegraphics[width=1.0\columnwidth]{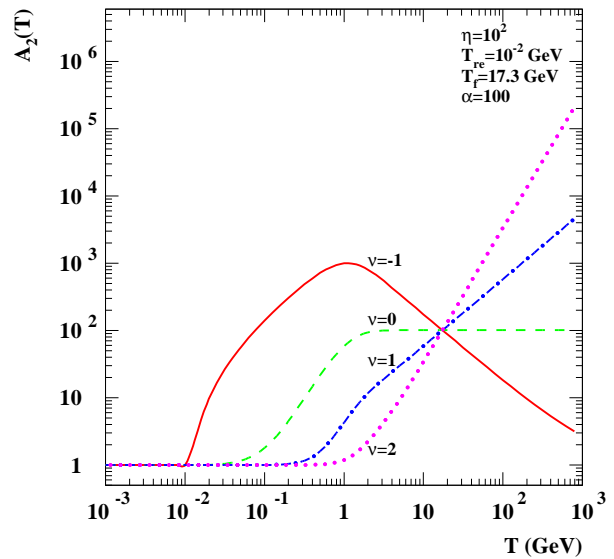}
\vspace{-20pt}
  \caption{Enhancement function $A_2(T)$ of Eq.~(\ref{A-2 def}) as a
  function of the temperature. Notice that time runs from right to
  left. The solid red line has $\nu = -1$, the dashed green line $\nu
  = 0 $, the dash-dotted blue $\nu = 1$ and the dotted purple line has
  $\nu = 2$. We have chosen $\alpha = 100$. All other parameters are
  identical to those of $A(T)$ in Fig.~\ref{fig:fnew01}, i.e.~$\eta =
  10^2$, $T_\textrm{re} = 10^{-2}$ GeV and $T_\textrm{f} = 17.3$ GeV.}
\label{fig:tva2}
\end{figure}

Now, let us go back to the situation where we have only a
temperature--independent thermal cross section but instead we modify
the enhancement of the Hubble rate. Let us assume, instead of the
enhancement $A(T)$ defined in Eq.~(\ref{A def}), the following
function:
\be
A_2(T)=1+\eta\,\left(\frac{T}{T_\textrm{f}}\right)^\nu\,
  \tanh^2\left(
 \frac{T-T_{\textrm{re}}}{\alpha T_{\textrm{re}}}\right)
\label{A-2 def}
\ee
The difference between $A(T)$ and $A_2(T)$ is that in the latter case
the \emph{tanh} is squared and in its argument we have introduced a
factor $\alpha$. This modification aims at obtaining (for large values
of $\alpha$) a ``re--entering'' into the standard regime less steep than in
the previous case. This is illustrated in fig.~\ref{fig:tva2}.

A comparison between the upper bound on $\eta$ in the case of a Hubble
enhancement $A(T)$ and $A_2(T)$ is shown again in
Fig.~\ref{fig:etamax04}. The central blue dotted line is for $A_2(T)$
with $\alpha = 100$, $T_\textrm{re} = 10^{-3}$ and $\nu = -1$. It can
be seen that this line practically falls on top of the limit (red
solid line) derived for $A(T)$ (and $T_\textrm{re} = 10^{-3}$, $\nu =
-1$). This result shows that the actual shape of the enhancement
function around ``re--entering'' is not of major importance as long as
the ``re--entering'' occurs long after the WIMP freeze out.

Let us finally consider how the upper bound on $\eta$ is affected by
the uncertainty in the astrophysical parameters of the propagation and
diffusion of the antiproton flux. We refer for this uncertainty to
Ref. \cite{pb}. Up to now, we have used a set of astrophysical
parameters which provides the median value of the antiproton flux
\cite{pb}. This set of parameters is the one which best fits cosmic
ray observations, mainly the B/C ratio \cite{PaperII}. We now show in
Fig.~\ref{fig:etamax04} the results obtained for the allowed range of
astrophysical parameters, which give a maximal and minimal antiproton
flux. The uppermost dotted curve in Fig.~\ref{fig:etamax04} shows the
result obtained using the ``minimal'' set of astrophysical parameters 
for the
propagation and diffusion, while the lowermost dotted curve shows the
result as calculated u\-sing the ``maximal'' set. These curves should
be compared to the central solid curve, which is our previous result
obtained with the best--fit parameters. For all the curves we have
$T_{\textrm{re}}=10^{-3}$ GeV and $\nu = -1$. We see that the
uncertainty in the propagation function causes an uncertainty of
almost one order of magnitude in both direction for the upper limit of
$\eta$, as expected from the discussion in Sec. \ref{sec:pbar} and
from the analytical approximations of Eq. (\ref{eq:etamax}).

From Fig. \ref{fig:etamax04} we also see that a different choice on
the astrophysical parameters may lead to a different lower limit on
the WIMP mass, as a consequence of the fact that lowering the bound on
the cross section (for the maximal set of astrophysical parameters) 
causes the upper bound on
$\eta$ to decrease. This can be seen from the plot of the $(\eta,\nu)$
plane in Fig.~\ref{fig:enmass07}. For a selection of masses this shows
the bound as calculated with the median propagation function. For a
given mass, the region to the left ({\em i.e.} at smaller $\eta$) is
allowed by the antiproton data as the cross section here is
lower. Thus, when we decrease the upper bound on the cross section,
{\em i.e.} increase the effect of the propagation parameters, all the
bounds in Fig.~\ref{fig:enmass07} will move to smaller $\eta$ (in 
particular the lower masses will move outside the plot). This is
consistent with the results of Ref. \cite{pbarlim}.

\section{Antiproton flux constraints on some specific models}
\label{sec:specific}

As we discussed in the previous sections, the maximum allowed
antiproton flux gives interesting model--independent constraints on
the maximum allowed enhancement of the expansion rate at dark matter
freeze out. Let us now discuss these constraints in connection to the
specific cosmological models we shortly introduced in Section
\ref{sec:cosm models ex}.

\subsection{Scalar-tensor theories}

As mentioned in section \ref{subsec:ST 1}, we do not have a general
analytical correspondence between the Hubble expansion rate in the
scalar--tensor model and our parametrization (\ref{A def}). However,
the numeric example of reference \cite{scalartensorrelic}, which
assumes a slowly varying scalar field, can be described by our
parametrization with $\nu = -0.82$ and $\eta =
A(\varphi(T_\textrm{f}))$, where $A(\varphi)$ is the Weyl factor,
which in this case gives the Hubble enhancement function $A(T)$.  From
figure \ref{fig:etamax03large} we get the upper bound:
\begin{equation}
A\left(\varphi(T_\textrm{f})\right)<\mathcal{O}(10^2) \qquad 
\textrm{for} \quad m_\chi\leq\mathcal{O}(500 \textrm{GeV}) 
\end{equation} 
and $T_\textrm{f}=\mathcal{O}(10 \textrm{GeV})$.  From figure
\ref{fig:avt3} we can also find the upper bound on $A(T) = A(\varphi)$
at earlier times. We see that $A(\varphi) < 10$ for
$T>\mathcal{O}(10^2 \textrm{GeV})$ and $m_\chi\leq\mathcal{O}(500
\textrm{GeV})$. It is interesting to observe that our results strongly
constrain Fig. 5 of \cite{scalartensorrelic} in which it could seem
that properly choosing the initial conditions for the scalar field,
each enhancement for the expansion rate is allowed. Antiproton data
limit this possibility.

\subsection{Kination}

In sec.~\ref{subsec:kination 1} we found that $\nu=1$ and $\eta =
\sqrt{\eta_{\Phi}} =
\sqrt{\rho_{\Phi}\left(T_\textrm{f}\right)/\rho_{r}\left(T_\textrm{f}\right)}$
can be used to describe the enhanced expansion rate during a kination
phase.

We know from Sec.~\ref{subsec:relic ratio} that $\eta$ is
approximately equal to the WIMP density enhancement. This is in
agreement with the conclusions of the earlier references on kination
and enhanced WIMP density; see {\em e.g.} Ref. \cite{Salati} and Ref.
\cite{profumoullio}. In Ref. \cite{profumoullio} they found that the
maximal density enhancement compatible with the BBN bounds is of the
order of $10^6$ (for WIMP masses smaller than 1 TeV), {\em i.e.}
$\eta_{\Phi}<10^{12}$. With our argument based on the maximum allowed
antiproton flux we can derive a much stronger constraint on this
scenario. The most conservative bound is found for large masses. For
$m_\chi< 1 \textrm{TeV}$ (and $\nu = 1$) we have
$\eta<\mathcal{O}(10^3)$ and so $\eta_{\Phi}<10^6$. While for light
WIMPs ($m_\chi<\mathcal{O}(10^2 \textrm{GeV})$) we have the much more
stringent bound $\eta<10$ and so $\eta_{\Phi}< 10^2$.

\subsection{Extra Dimensions}

As mentioned in section \ref{subsec:extra dim 1}, the RSII model
\cite{Randall} gives rise to an enhanced expansion rate which can be
described by our parametrization with the values of the parameters
$\nu=2$ and $\eta=\sqrt{\rho_r\left(T_\textrm{f}\right) /
(2\lambda)}$. Where $\lambda$ is the tension on the brane and $\rho_r$
is the radiation ener\-gy density. The upper bound on $\eta$ therefore
gives us a lower bound on the brane tension. The upper bound on $\eta$
for $\nu = 2$ is shown in Fig.~\ref{fig:etamax03large}. For
$\eta<\mathcal{O}(10^2)$, which is true for WIMP masses
$m_\chi<\mathcal{O}(10^2 \textrm{GeV})$, we get the most stringent
bound. With $T_\textrm{f}\simeq 5 \ \textrm{GeV}$ and
$g_{\textrm{{eff}}}(T_\textrm{f})\simeq 9^2$ we find the bound
$\lambda>\mathcal{O}(1 \textrm{GeV}^4)$. A more conservative bound is
found for higher masses. For $m_\chi = \mathcal{O}(500 \
\textrm{GeV})$ we have $\eta<\mathcal{O}(10^4)$ and
$T_\textrm{f}\simeq 20 \ \textrm{GeV}$, leading to
$\lambda>\mathcal{O}(10^{-2} \textrm{GeV}^4)$. Finally, using the
relation \cite{Durrer}
\begin{equation}
\lambda=\frac{3}{4\pi} \frac{M_5^6}{M_{pl}^2}
\end{equation}
where $M_{pl}$ is the four dimensional Plank mass and $M_5$ the five
dimensional one, we can derive a lower bound on $M_5$. For the most
stringent bound on $\lambda$ we get $M_5 > 3 \cdot 10^3 \
\textrm{TeV}$, while the more conservative bound just give a small
change, $M_5 > 10^3 \ \textrm{TeV}$. We observe that the BBN bound is
$\lambda>1 \ \textrm{MeV}^4$ and so $M_5 > 30 \ \textrm{TeV}$,
\cite{Durrer}. Therefore in this context, our analysis based on the
antiproton flux is much more stringent than the one coming form the
BBN constraint.

\section{Conclusion}
\label{sec:conclusion}

The expansion rate of the Universe in the era before BBN is predicted
to be different from the one of standard FRW cosmology in a host of
cosmological models which are derived either from modification of General
Relativity, or from brane physics, or from attempts to provide a
consistent explanation to dark energy. In the present paper we
concentrated on models which predict an enhancement of the expansion
rate, like {\em e.g.} those discussed in Ref. \cite{scalartensorrelic}
where the dark energy problem finds a solution in a scalar--tensor
theory of gravity with an exponential run--away behaviour of a scalar
field, in kination models like those of
Ref. \cite{Salati,Rosati,profumoullio,Pallis}, in D--brane models like
the RSII model of Ref. \cite{Randall} or models of modified cosmology
like those in Ref. \cite{Barrow,kamionkowskiturner}. These
enhancements may be sizeable, without evading the post--BBN bounds
(light primordial element production, CMB anisotropies, gravitational
tests) which may be applied to these cosmolo\-gi\-cal models: the
enhancements may reach up to four \cite{scalartensorrelic} or even six
\cite{Salati,profumoullio} orders of magnitude. However, an imprint of
this enhancement may be left on the properties of the particles
responsible of the dark matter. This happens if the enhancement of the
expansion rate occurs close to the time when the dark matter particles
decouples from the thermal bath: this possibility relies on the fact
that the cosmological bound on the amount of dark matter can be
fulfilled by particles with larger annihilation cross section, as
compared to standard cosmology \cite{scalartensorrelic,Salati}, due to
an anticipated freeze--out. In this situations, indirect detection
signals are typically enhanced, as long as the thermal cross section
responsible of the relic abundance is close to the one which determines 
the indirect detection signals (like {\em e.g.} when the typical
temperature expansion of the thermal cross section is applicable).

Cosmic antiprotons produced by WIMPs in the Galaxy are currently the
best option for constraining particle dark matter annihilation cross
sections \cite{pbarlim,pb}, since they are not strongly dependent on
assumptions on the dark matter density profile, like instead is the
case of the gamma--ray signal. Also, cosmic antiprotons produced by 
WIMPs are naturally predicted to be
at the level of the background and of the data, when the WIMP accounts 
for the CDM content of the Universe, and without invoking high
degrees of over--densities in the galactic neighbourhood
\cite{pb}. Since the experi\-mental data are in good agreement with the
expected background \cite{PaperII,revue}, not much room is left for an
exotic component and therefore antiprotons may be used to set
constraints \cite{pb,pbarlim}.

In this paper we exploited the effect induced on the antiproton signal
by an increased expansion rate, to set bounds on the maximal
enhancement of the expansion rate which can be allowed in the early
Universe in a pre--BBN epoch \cite{note on reduction}. These limits apply 
to an evolutionary phase close to
dark matter decoupling ({\em i.e.}  at temperatures of the Universe in
the range of 1--30 GeV, for dark matter of about 10--1000 GeV mass)
and are subject to the (quite reasonable) hypothesis that dark matter
is a thermal relic. The bounds we have obtained are quite strong,
especially for light dark matter. Our main result is summarized by 
Fig.~\ref{fig:etamax03large}. For WIMP masses lighter than 100
GeV, the bound on the Hubble--rate enhancement ranges from a factor of
a few to a factor of 30, depending on the actual cosmological model,
while for a mass of 500 GeV the bound falls in the range
50--500. These bounds are affected by the still large astrophysical
uncertainties in the calculation of the antiproton signal \cite{pb},
which reflects in a factor of 5 more stringent, or a factor of 10 more
loose bounds. Nevertheless, the limits we obtain are much more
stringent than the possible enhancement of these cosmological models.

A caveat is in order here: whenever the effective thermal annihilation
cross section in the early Universe (responsible for the relic
abundance) is sizably larger than the one in the galactic halo
(which determines the antiproton signal) the bounds we determine are
necessarily loosened. This occurs when coannihilation effects are
relevant in determining the relic abundance. In this case the
annihilation cross section in the Galaxy is typically much smaller
than the effective one which sets the decoupling: in this case our
limits still apply, but they are less constraining, approximately
by the ratio of the two cross sections. We nevertheless notice that
the occurrence of coannihilation is usually an accidental feature.
Another case where our limits are less severe is when WIMP
annihilation occurs mostly into leptons: in this case antiprotons are
not produced and our bounds are loosened by a factor given by the
branching ratio into non-leptonic final states. In all other situations,
which represent the most typical cases, the limits we quote are
actually necessarily present.

Finally, we applied our bounds to set constraints on some specific
models. In the case of the scalar--tensor dark energy models of
Ref. \cite{scalartensorrelic}, we constrain the enhancement to be less
than a factor 100 for dark matter lighter than 500 GeV. In the case of
kination models, the maximal enhancement is 1000 for WIMPs lighter
than 1 TeV, while for dark matter lighter than 100 GeV the enhancement
must be below 10. Finally, for the Randall--Sundrum D--brane model of
Ref. \cite{Randall}, our limit may be used to set bounds on the
tension on the brane: for WIMPs lighter than 100 GeV, the string
tension must be larger than 1 GeV$^4$; for heavier dark matter, the
bound is 10$^{-2}$ GeV$^4$. These limits reflect in a lower bound of
about 10$^3$ TeV on the 5--dimensional Planck mass and are much more
severe than the usual constraints derived from BBN physics.

%%%%%%%%%%%%%%%%%%%%%%%%%%%%%%%%%%%%%%%%%%%%%%%%%%%%%%%%%%%%%%%%%%%%%%%%%%%%%%%%%%%%%%
\appendix

\section{Approximate Boltzmann--equation solutions for the modified Hubble rate}
\label{app:analytic}

The Boltzmann equation Eq. (\ref{eq:boltzmann0}) may be easily solved
for a modified Hubble rate as defined in Eq. (\ref{A def}) by suitable
approximations. We first make explicit the temperature (or
$x=m_\chi/T$) dependence in Eq. (\ref{eq:boltzmann0}) by rewriting the
equation in the usual form:
\begin{equation}
\frac{dY}{dx} = -\sqrt{\frac{\pi}{45 G}}
\frac{h_{\textrm{eff}}(x)}{\sqrt{A^2(x)g_{\textrm{eff}}(x)}}
\frac{m_\chi}{x^2}\langle\sigma_{\textrm{ann}}v\rangle (Y^2-Y^2_\textrm{eq})
\label{eq:boltzmann1}
\end{equation}
where $G$ is the Newton constant. The standard cosmo\-lo\-gy case 
is re\-co\-ve\-red for
$A(T)=1$.

We first define the temperature of particle freeze--out as
the temperature $T_f$ when:
\begin{equation}
Y(x_{\rm f}) = (1+c) Y_{\rm eq} (x_{\rm f})
\label{eq:freeze condition}
\end{equation}
with $c$ a constant of order 1 and $Y_{\rm eq}(x) = 45\,\times\, 2^{-5/2}
\pi^{-7/2}\, g\, h^{-1}_{\rm eff}(x)\, x^{-3/2}\, \exp(-x)$ is the equilibrium
abundance of a non--relativistic particle with internal degree of
freedom $g$. Condition (\ref{eq:freeze condition}) and Eq. (\ref{eq:boltzmann1})
lead to the following implicit expression for the freeze--out
temperature:
\begin{equation}
x_\textrm{f} = \ln\left[\frac{2c}{1+c} 0.038 \,m_\textrm{pl}\,g\,m_\chi 
\frac{\langle\sigma_{\textrm{ann}}v\rangle_{x_\textrm{f}}}
{x_\textrm{f}^{1/2}\,A(x_{\rm f})\,g^{1/2}_{\textrm{eff}}(x_\textrm{f})}
\right]
\label{eq:freeze}
\end{equation}
where $m_{\rm pl}$ is the Planck mass. For definiteness, we use $c=1$
throughout the paper. The standard freeze--out temperature is
obtained when $A(T)=1$ around the time of particle decoupling. The
models we are considering, where $A(T)>1$ at that epoch, imply a
smaller $x_{\rm f}$, {\em i.e.} a larger freeze--out temperature $T_f$.

The current value $Y_0$ of the comoving abundance is then obtained by
integrating Eq. (\ref{eq:boltzmann1}) from $x_{\rm f}$ down to $x_0
\rightarrow \infty$, neglecting the WIMP production term:
\begin{equation}
\frac{dY}{dx} = -\sqrt{\frac{\pi}{45G}}
\frac{h_{\textrm{eff}}(x)}{\sqrt{A^2(x)g_{\textrm{eff}}(x)}}
\frac{m_\chi}{x^2}\langle\sigma_{\textrm{ann}}v\rangle Y^2
\end{equation}
which gives the solution:
\begin{equation}
\frac{1}{Y_0} = \frac{1}{Y_{\rm eq}(x_{\rm f})} + B\, m_\chi \, \int_{x_{\rm f}}^\infty
\frac{{\cal G}(x)\,\langle\sigma_{\textrm{ann}}v\rangle}{A(x)\, x^2}\, dx
\label{eq:Y0}
\end{equation}
where $B=(\pi/45 G)^{1/2}$ and ${\cal G}(x) = h_{\textrm{eff}}(x)/
\sqrt{g_{\textrm{eff}}(x)}$. The WIMP relic abundance is then obtained
by means of Eq. (\ref{eq:relic}).

For large enhancements of the Hubble rate around the WIMP freeze--out,
{\em i.e.} for large values of the parameter $\eta$, we may approximate
$A(T)$ as a power--law function with exponent $\nu$. If $\nu<0$, we
may assume this behaviour all the way down to the re--entering
temperature $x_{\rm re}$:
\begin{equation}
A(T) \simeq \eta \left(\frac{T}{T_f}\right)^\nu = (\eta\,x_{\rm f}^\nu)\,x^{-\nu}
\end{equation}
When $\nu>0$ the above approximation may not be used up to $x_{\rm
re}$, because $A(x)$ approaches unity as the Universe cools down (see
Fig.~\ref{fig:fnew01}) and therefore we cannot neglect the 1 in
Eq.~(\ref{A def}). We define as $x_1$ the temperature when
$(\eta\,x_{\rm f}^\nu)\,x^{-\nu} \simeq 1$, {\em i.e.} $x_1 = x_{\rm f}
\eta^{1/\nu}$. In this case we approximate:
\begin{equation}
A(T) \simeq \left \{ 
  \begin{array}{ll}
    (\eta\,x_{\rm f}^\nu)\,x^{-\nu} & \quad\quad\quad x \lsim x_1\\
    1                     & \\
  \end{array}
\right.
\end{equation}
Below $x_{\rm re}$, in any case $A(T)=1$ by definition.

With these approximations, the solution of Eq.~(\ref{eq:Y0}) for a
temperature--independent annihilation cross section
$\langle\sigma_{\textrm{ann}}v\rangle = a$ is:
\begin{eqnarray}
\label{eq:sol1}
\frac{1}{Y_0} &=& \frac{1}{Y_{\rm eq}(x_{\rm f})} \\
        &+& B\, m_\chi \, a \, {\cal G}(x_{\rm f})\, 
        \left [
            {\cal I}(x_{\rm f},\hat x,\eta,\nu) + r_G\, \frac{1}{\hat x} 
        \right ] \nonumber
\end{eqnarray}
where $\hat x = \min[x_1,x_{\rm re}]$ and $r_G={\cal G}(\hat x)/{\cal
G}(x_{\rm f})$ is typically around 0.5--1 for WIMPs masses in the
range GeV--TeV. For large enhancements ($\eta \gg 1$) and a low
re--entering temperature, typically: $\hat x = x_1$ for $\nu>0$ and
$\hat x = x_{\rm re}$ for $\nu<0$. In Eq. (\ref{eq:sol1}) we have
assumed a slowly--varying ${\cal G}(x)$ in the integrals.

In our discussion in the text, typically the relevant term in
Eq.~(\ref{eq:sol1}) is the one which contains the function ${\cal
I}(x_{\rm f},\hat x,\eta,\nu)$: the other two terms may be usually neglected to
a few percent level of approximation. This relevant function is
easily found to be:
\begin{equation}
{\cal I} = 
   \frac{1}{\nu-1}\,\frac{1}{x_{\rm f}\,\eta^{1/\nu}}
\label{eq:ifunction1}
\end{equation}
for $\nu > 1$, then
\begin{equation}
{\cal I} = 
   \frac{1}{1-\nu}\,\frac{1}{x_{\rm f}\,\eta}
\label{eq:ifunction2}
\end{equation}
for $\nu < 1$ and finally
\begin{equation}
{\cal I} =
   \frac{1}{x_{\rm f}}\,\frac{\ln(\eta)}{\eta}
\label{eq:ifunction3}
\end{equation}
for $\nu = 1$ (kination).

\acknowledgments 
We acknowledge Research Grants funded jointly by the Italian Ministero
dell'Istruzione, dell'Universit\`a e della Ricerca (MIUR), by the
University of Torino and by the Istituto Nazionale di Fisica Nucleare
(INFN) within the {\sl Astroparticle Physics Project}.  R. Catena
acknowledges a Research Grant funded by the VIPAC Institute.

%%%%%%%%%%%%%%%%%%%%%%%%%%%%%%%%%%%%%%%%%%%%%%%%%%%%%%%%%%%%%%%%%%%%


\begin{thebibliography}{99}

\bibitem{omegalimits} D.N. Spergel {\em et al.} (WMAP Collaboration),
[arXiv:astro--ph/0603449] (submitted to Astrophys. J.)

\bibitem{note on reduction} A reduction of the expansion rate is also 
possible; see {\em e.g.}~Ref.~\cite{reduction}. We are not going to discuss 
these 
models here, since when the Hubble rate is rduced, the antiproton bounds 
derived in the paper does not lead to relevant constraints.

\bibitem{reduction} R.~Catena, M.~Pietroni and L.~Scarabello, Phys.~Rev.~D 
{\bf 70}, 103526 (2004) [arXiv:astro-ph/0407646]; C.~de Rham, T.~Shiromizu 
and A.~J.~Tolley, [arXiv:gr-qc/0604071].  

\bibitem{scalartensorrelic} R.~Catena, N.~Fornengo, A.~Masiero,
M.~Pietroni and F.~Rosati, Phys.\ Rev.\ D {\bf 70}, 063519 (2004)
[arXiv:astro-ph/0403614].

\bibitem{Salati}
  P.~Salati,
  %``Quintessence and the relic density of neutralinos,''
  Phys.\ Lett.\ B {\bf 571}, 121 (2003).

\bibitem{Rosati}
  F.~Rosati, Phys.~Lett.~B {\bf 570}, 5 (2003) [arXiv:hep-ph/0309124].

\bibitem{profumoullio} S.~Profumo and P.~Ullio, JCAP {\bf 0311},
006 (2003) [arXiv:hep-ph/0309220]

\bibitem{Pallis}
  C.~Pallis, JCAP {\bf 0510}, 015 (2005) [arXiv:hep-ph/0503080].


\bibitem{Barrow} 
  J.~D.~Barrow, Nucl.~Phys.~B {\bf 208}, 501 (1982).

\bibitem{kamionkowskiturner}
  M.~Kamionkowski and M.~S.~Turner, Phys.~Rev.~D {\bf 42}, 3310 (1990).

\bibitem{Randall}
  L.~Randall and R.~Sundrum,
  %``An alternative to compactification,''
  Phys.\ Rev.\ Lett.\  {\bf 83}, 4690 (1999).

\bibitem{pb} F.~Donato, N.~Fornengo, D.~Maurin, P.~Salati and
  R.~Taillet, Phys.\ Rev.\ {\bf D69}, 063501 (2004)
  [arXiv:astro-ph/0306207].

\bibitem{pbarlim} A.~Bottino, F.~Donato, N.~Fornengo and P.~Salati,
  Phys.\ Rev.\ {\bf D72}, 083518 (2005).

\bibitem{Cyburt}
  R.~H.~Cyburt, B.~D.~Fields, K.~A.~Olive and E.~Skillman,
  %``New BBN limits on physics beyond the standard model from He-4,''
  Astropart.\ Phys.\  {\bf 23}, 313 (2005)
  [arXiv:astro-ph/0408033].

\bibitem{Wang}
  Y.~Wang and P.~Mukherjee,
  %``Robust Dark Energy Constraints from Supernovae, Galaxy Clustering, and
  %Three-Year Wilkinson Microwave Anisotropy Probe Observations,''
  [arXiv:astro-ph/0604051].

\bibitem{Zhao}
  G.~B.~Zhao, J.~Q.~Xia, B.~Feng and X.~Zhang,
  %``Probing dynamics of dark energy with supernova, galaxy clustering and the
  %three-year Wilkinson Microwave Anisotropy Probe (WMAP) observations,''
  [arXiv:astro-ph/0603621].

\bibitem{Albert}
  J.~Albert {\it et al.}  [SNAP Collaboration],
  %``Probing Dark Energy via Weak Gravitational Lensing with the SuperNova
  %Acceleration Probe (SNAP),''
  [arXiv:astro-ph/0507460].

\bibitem{Cassini}
  B.~Bertotti, L.~Iess and P.~Tortora,
  %``A test of general relativity using radio links with the Cassini
  %spacecraft,''
  Nature {\bf 425}, 374 (2003).

\bibitem{Damour}
T. ~Damour, [arXiv:gr-qc/9606079], lectures given at Les Houches 1992, 
SUSY05 and Corfu' 1995.

\bibitem{Esposito}
  G.~Esposito-Farese and D.~Polarski,
%   ``Scalar-tensor gravity in an accelerating universe,''
  Phys.\ Rev.\ D {\bf 63}, 063504 (2001)
  [arXiv:gr-qc/0009034].
 
\bibitem{Dicke}
 R.~H.~Dicke,
%   ``Mach's Principle And Invariance Under Transformation Of Units,''
  Phys.\ Rev.\  {\bf 125}, 2163 (1962).

\bibitem{Pietroni}
  R.~Catena, M.~Pietroni and L.~Scarabello,
  %``Einstein and Jordan reconciled: a frame-invariant approach to 
  %scalar-tensor cosmology,''
  [arXiv:astro-ph/0604492].


\bibitem{Steinhardt} P.~J.~Steinhardt, L.~M.~Wang and I.~Zlatev, Phys.~Rev.~D 
{\bf 59}, 123504 (1999) [arXiv:astro-ph/9812313].


\bibitem{Shiromizu}
  T.~Shiromizu, K.~I.~Maeda and M.~Sasaki,
  %``The Einstein equations on the 3-brane world,''
  Phys.\ Rev.\ D {\bf 62}, 024012 (2000).

\bibitem{Durrer}
  R.~Durrer,
  %``Braneworlds,''
  AIP Conf.\ Proc.\  {\bf 782}, 202 (2005)
  [arXiv:hep-th/0507006].

\bibitem{PaperII}
F.~Donato, D.~Maurin, P.~Salati, A.~Barrau, G.~Boudoul, and 
R.~Taillet, Astrophys. J. {\bf 563}, 172 (2001).

\bibitem{revue}
D. Maurin, R. Taillet, F. Donato, P. Salati, A. Barrau, and G. Boudoul,
Research Signposts (2004), ``Recent Research Developments in Astronomy and
Astrophysics'', v.2, p.193 [arXiv:astro-ph/0212111].

\bibitem{PaperI}
D.~Maurin, F.~Donato, R.~Taillet, and P.~Salati,
Astrophys. J. {\bf 555}, 585 (2001).

\bibitem{bess95-97} S. Orito, {\etal} ({\sc BESS} Collaboration),
Phys. Rev. Lett. {\bf 84}, 1078 (2000).

\bibitem{bess98} T. Maeno, {\etal} ({\sc BESS} Collaboration),
Astropart. Phys. {\bf 16}, 121 (2001).

\bibitem{ams98} M Aguilar, {\etal} ({\sc AMS} Collaboration),
Phys. Rep. {\bf 366}, 331 (2002).

\bibitem{caprice} M. Boezio, {\etal} ({\sc CAPRICE} Collaboration),
Astrophys. J. {\bf 561}, 787 (2001).

\bibitem{boost} J.~Lavalle, J.~Pochon, P.~Salati and R.~Taillet, [arXiv:astro-ph/0603796]. 

\bibitem{dbar} F. Donato, N. Fornengo, P. Salati, Phys.\ Rev.\ {\bf
D62}, 043003 (2000).

\bibitem{gaps} K. Mori {\em et al.}, Astrophys. J. {\bf 566}, 604 
(2002) [arXiv:astro-ph/0109463].

\bibitem{ams} M.~Aguilar {\em et al.} (AMS Collaboration),
Phys. Rep. {\bf 366/6}, 331 (2002).

\bibitem{sweden} P.~Ullio, [arXiv:astro-ph/9904086].

\bibitem{coannihil} 
S.~Mizuta and M.~Yamaguchi,
Phys.\ Lett.\ B {\bf 298}, 120 (1993); J.~Edsjo and P.~Gondolo,
Phys.\ Rev.\ D {\bf 56}, 1879 (1997)
[arXiv:hep-ph/9704361]; 
J.~R.~Ellis, T.~Falk and K.~A.~Olive, Phys.\ Lett.\ B {\bf 444}, 367 (1998)
[arXiv:hep-ph/9810360].

\bibitem{pole1} P.~Bin{\'e}truy, G.~Girardi and P.~Salati, Nucl.~Phys. B 
{\bf 237}, 285 (1984) 
; K.~Griest and D.~Seckel, Phys.\ Rev.\ D {\bf 43}, 3191
(1991)

\bibitem{pole2} P. Gondolo, and G. Gelmini, Nucl.~Phys.~B {\bf 360}, 145 (1991).

\end{thebibliography}
\end{document}